\documentclass[oneside]{jfm}
\usepackage{graphicx}
\usepackage{newtxtext}
\usepackage{newtxmath,amsmath}
\usepackage[makeroom]{cancel}
\usepackage{bm, comment}
\usepackage{natbib}
\usepackage{hyperref}
\hypersetup{
    colorlinks = true,
    urlcolor   = blue,
    citecolor  = blue,
}
\usepackage{color}
\usepackage{multirow}
\usepackage[normalem]{ulem} 

\newcommand{\RomanNumeralCaps}[1]
\linenumbers

\graphicspath{{./figures/}}

\title[Correlation time of velocity and kinetic helicity]
{Correlation times of velocity and kinetic helicity fluctuations in nonhelical hydrodynamic turbulence}

\author[Zhou \& Jingade]
{Hongzhe Zhou\aff{1,2}
\corresp{\email{hongzhe.zhou@sjtu.edu.cn}},
Naveen Jingade\aff{3}
\corresp{\email{naveen.jingade@iiap.res.in}}}

\affiliation{\aff{1}Tsung-Dao Lee Institute, Shanghai Jiao Tong University, Shanghai, 201210, China
\aff{2}Nordita, KTH Royal Institute of Technology and Stockholm University,
Hannes Alfv\'ens v\"ag 12, SE-106 91 Stockholm, Sweden
\aff{3}Indian Institute of Astrophysics, Koramangala, Bengaluru-560034\\
}

\begin{document}
\newcommand{\beq}{\begin{equation}}
\newcommand{\eeq}{\end{equation}}

\newcommand{\benum}{\begin{enumerate}}
\newcommand{\eenum}{\end{enumerate}}

\def\apj{ApJ}
\def\apjs{ApJS}
\def\aap{A \& Ap}
\def\pasp{PASP}
\def\pasj{PASJ}
\def\ssr{Sp Sci. Rev.}
\def\solphys{Solar Physics}
\def\mA{\mathcal A}
\def\araa{ARAA}
\def\nat{Nature}
\def\mnras{MNRAS}
\def\apj{ApJ}
\def\apjl{ApJL}
\def\aap{A\&A}
\def\apss{Ap\& Spac. Sup.}
\def\jcap{Journal of Cosmology and Astroparticle Physics}
\def\physrep{Physics Reports}
\def\pre{Physical Review E}
\def\physscr{Physica Scripta}
\def\aj{AJ}
\def\prl{Phys. Rev. Lett.}
\def\pnas{Proceedings of the National Academy of Science}
\def\qjras{Quarterly Journal of the Royal Astronomical Society}

\newcommand{\titlealpha}{\texorpdfstring{$\bm\alpha$}{p}}
\definecolor{dred}{rgb}{.9, .15,.1}
\definecolor{brown}{rgb}{0.5,0.2,0.2}
\definecolor{dgreen}{rgb}{0.15,0.4,0.15}
\definecolor{dblue}{rgb}{0.0,0.0,0.6}

\def\bar{\overline}
\newcommand{\abra}[1]{\left\langle{#1}\right\rangle}
\newcommand\remark[1]{{\color{red}[{\it{#1}}]}}
\newcommand{\hz}[1]{{\color{magenta}[HZ: {#1}]}}
\newcommand{\stkout}[1]{\ifmmode\text{\sout{\ensuremath{#1}}}\else\sout{#1}\fi}

\newcommand{\bmA}{{\bm A}}
\newcommand{\bma}{{\bm a}}
\newcommand{\bmB}{{\bm B}}
\newcommand{\bmb}{{\bm b}}
\newcommand{\bmU}{{\bm U}}
\def\bmu{\bm{u}}
\newcommand{\bmJ}{{\bm J}}
\newcommand{\bmj}{{\bm j}}
\newcommand{\bmk}{{\bfk}}
\newcommand{\bmK}{{\bfK}}
\newcommand{\bmx}{{\bfx}}
\newcommand{\bmE}{{\bm{E}}}
\newcommand{\bme}{{\bm{e}}}

\newcommand{\bfa}{\mbox{\boldmath $a$}}
\newcommand{\bfb}{\mbox{\boldmath $b$}}
\newcommand{\bfc}{\mbox{\boldmath $c$}}
\newcommand{\bfe}{\mbox{\boldmath $e$}}
\newcommand{\bff}{\mbox{\boldmath $f$}}
\newcommand{\bfg}{\mbox{\boldmath $g$}}
\newcommand{\bfh}{\mbox{\boldmath $h$}}
\newcommand{\bfk}{\mbox{\boldmath $k$}}
\newcommand{\bfn}{\mbox{\boldmath $n$}}
\newcommand{\bfq}{\mbox{\boldmath $q$}}
\newcommand{\bfu}{\mbox{\boldmath $u$}}
\newcommand{\bfv}{\mbox{\boldmath $v$}}
\newcommand{\bfx}{\mbox{\boldmath $x$}}
\newcommand{\bfy}{\mbox{\boldmath $y$}}
\newcommand{\bfr}{\mbox{\boldmath $r$}}
\newcommand{\bfA}{\mbox{\boldmath $A$}}
\newcommand{\bfB}{\mbox{\boldmath $B$}}
\newcommand{\bfC}{\mbox{\boldmath $C$}}
\newcommand{\bfG}{\mbox{\boldmath $G$}}
\newcommand{\bfJ}{\mbox{\boldmath $J$}}
\newcommand{\bfE}{\mbox{\boldmath $E$}}
\newcommand{\bfF}{\mbox{\boldmath $F$}}
\newcommand{\bfH}{\mbox{\boldmath $H$}}
\newcommand{\bfK}{\mbox{\boldmath $K$}}
\newcommand{\bfM}{\mbox{\boldmath $M$}}
\newcommand{\bfR}{\mbox{\boldmath $R$}}
\newcommand{\bfU}{\mbox{\boldmath $U$}}
\newcommand{\bfV}{\mbox{\boldmath $V$}}
\newcommand{\bfW}{\mbox{\boldmath $W$}}
\newcommand{\bfX}{\mbox{\boldmath $X$}}
\newcommand{\bfY}{\mbox{\boldmath $Y$}}
\newcommand{\bfQ}{\mbox{\boldmath $Q$}}
\newcommand{\bfxi}{\mbox{\boldmath $\xi$}}
\newcommand{\ex}{\mbox{{\boldmath $e$}}_{x}}
\newcommand{\ey}{\mbox{{\boldmath $e$}}_{y}}
\newcommand{\ez}{\mbox{{\boldmath $e$}}_{z}}
\newcommand{\vareps}{\mbox{{\boldmath $\varepsilon$}}}
\newcommand{\ec}{\mbox{{\boldmath $e$}}_{c}}
\newcommand{\eb}{\mbox{{\boldmath $e$}}_{b}}
\newcommand{\ek}{\mbox{{\boldmath $e$}}_{k}}
\newcommand{\cross}{\mbox{\boldmath $\times$}}
\newcommand{\cendot}{\mbox{\boldmath $\cdot\,$}}
\newcommand{\pr}{{\rm Pr}}
\newcommand{\pe}{{\rm Pe}}
\newcommand{\bsy}{\boldsymbol}
\newcommand{\ud}{\mathrm{d}}
\newcommand{\Eq}[1]{Equation~(\ref{#1})}
\newcommand{\Eqs}[2]{Equations~(\ref{#1}) and~(\ref{#2})}
\newcommand{\EEqs}[2]{Equations~(\ref{#1}) and~(\ref{#2})}
\newcommand{\Eqss}[2]{Equations~(\ref{#1})--(\ref{#2})}
\newcommand{\Fig}[1]{Figure~\ref{#1}}
\newcommand{\Figs}[2]{Figures~\ref{#1} and \ref{#2}}
\newcommand{\Sec}[1]{Section~(\ref{#1})}

\newcommand{\emf}{\mathcal{E}}
\newcommand{\bmemf}{{\bm{\emf}}}
\newcommand{\del}{{\bm\nabla}}
\newcommand{\nuM}{{\nu_\text{M}}}
\newcommand{\alphaK}{{\alpha_\text{K}}}
\newcommand{\alphaM}{{\alpha_\text{M}}}
\newcommand{\lL}{{l_\text{L}}}
\newcommand{\led}{{l_\text{ed}}}
\newcommand{\taued}{{\tau_\text{ed}}}

\newcommand{\ReN}{\text{Re}}
\newcommand{\ReM}{\text{Rm}}
\newcommand{\tcu}{t_{\text{c}}^u}
\newcommand{\tch}{t_{\text{c}}^h}
\newcommand{\tc}{\tau}
\newcommand{\ttc}{\tilde\tau}
\newcommand{\urms}{u_\text{rms}}
\newcommand{\Hrms}{H_\text{rms}}
\newcommand{\kf}{k_\text{f}}
\newcommand{\knu}{k_\nu}
\newcommand{\Ro}{\text{Ro}}
\newcommand{\Co}{\text{Co}}
\newcommand{\Cokf}{\text{Co}_\text{f}}
\newcommand{\Sh}{\text{Sh}}
\newcommand{\dtcor}{\Delta}
\newcommand{\corr}[1]{\text{Corr}\left[{#1}\right]}
\newcommand{\Omegak}{{\Omega_{\bm k}}}
\newcommand{\kr}{k_r}
\newcommand{\ktheta}{k_\theta}
\newcommand{\kphi}{k_\phi}
\newcommand{\ted}{\tau_\text{ed}}

\newcommand{\dt}{\text{d}t}
\newcommand{\gstar}{\gamma^*}
\newcommand{\kzstar}{k^*}
\newcommand{\kzmax}{k_{\text{max}}}

\newcommand{\Corr}{\text{Corr}}
\newcommand{\Cu}[1]{C_{u}^\text{{#1}}}
\newcommand{\Ctu}[1]{C_{\tilde u}^\text{{#1}}}
\newcommand{\Cg}[1]{C_{g}^\text{{#1}}}
\newcommand{\Ctg}[1]{C_{\tilde g}^\text{{#1}}}

\newcommand{\tautu}[1]{\tau_{\tilde u}^\text{{#1}}}
\newcommand{\tautg}[1]{\tau_{\tilde g}^\text{{#1}}}
\newcommand{\taug}{\tau_g}
\newcommand{\zt}[1]{\zeta^\text{{#1}}}

\newcommand{\kze}{k_\text{ze}}
\newcommand{\Ma}{\text{Ma}}

\def\qqq#1{{\color{brown}$\ \ \ $\bf Comments: #1}}
\def\blue{\textcolor{dblue}}
\def\red{\textcolor{dred}}

\newcommand{\xcancelto}[2]{\mathrlap{\raisebox{0.5ex}{\color{red}{\rule{4ex}{0.5pt}}}}#1\rightarrow #2}

\newcommand{\Cfull}{\cal T}

\newcommand{\revise}[2]{{{\color{red}\stkout{#1}}}{ }{\color{dgreen}{#2}}}
\renewcommand{\revise}[2]{{\color{black}{#2}}}

\maketitle

\begin{abstract}
Nonhelical turbulence within a linear shear flow has demonstrated efficient amplification of large-scale magnetic fields in numerical simulations,
but its precise mechanism remains elusive.
The incoherent $\alpha$ mechanism proposes that a zero-mean fluctuating transport coefficient $\alpha$ (linked to kinetic helicity) in the shear flow is a candidate driver.
Previous renovating-flow models have proposed that the correlation time of helicity fluctuations must be sufficiently extended to overcome turbulent magnetic diffusivity, yet only empirical validation of this concept \revise{remains outstanding}{has been obtained}.
In this study, we conduct direct numerical simulations of weakly compressible nonhelical hydrodynamic turbulence.
We scrutinize the correlation times of velocity and kinetic helicity fluctuations in distinct flow configurations, including rotation, shearing, and Keplerian flows, as well as the shearing burgulence counterpart.
Our findings indicate that rotation contributes to a prolonged correlation time of helicity compared to velocity, particularly notable in auto-correlations of both volume-averaged quantities and individual Fourier modes due to the formation of large-scale vortices.
In contrast, moderate shear strength does not exhibit significant scale separation, with shear flows elongating vortices in the shear direction.
Shearing burgulence, characterized by shorter helicity correlation times, appears less conducive to hosting the incoherent $\alpha$ effect.
Notably, at modest shear rates, only Keplerian flows exhibit sufficiently coherent helicity fluctuations, in contrast to shearing flows.
However, the relative strength of helicity fluctuations compared to turbulent diffusivity is significantly lower, raising doubts about the viability of the incoherent $\alpha$ effect as a potential dynamo driver in the subsonic flows examined in this study.
\end{abstract}

\begin{keywords}
Authors should not enter keywords on the manuscript, as these must be chosen by the author during the online submission process and will then be added during the typesetting process (see \href{https://www.cambridge.org/core/journals/journal-of-fluid-mechanics/information/list-of-keywords}{Keyword PDF} for the full list).  Other classifications will be added at the same time.
\end{keywords}

\section{Introduction}

Magnetic fields in astrophysical systems, encompassing celestial bodies such as stars, galaxies, and accretion disks, necessitate sustained mechanisms for their persistence against the effects of microphysical dissipation, a role aptly fulfilled by dynamos.
The study of magnetic field evolution, especially on scales exceeding the outer scale of turbulence, is conventionally approached through the prism of large-scale or mean-field dynamo theories \citep{SKR1966}.
A fundamental driver in mean-field dynamos is the net kinetic helicity intrinsic to the flow \citep{Parker1955} which naturally emerges in the presence of rotation and density stratification, commonly encapsulated in the $\alpha$ effect.
In this context, the mean-field dynamo coefficient $\alpha$ intricately depends on the product of kinetic helicity and the eddy turnover time.

The last two decades have witnessed the recognition of a novel class of mean-field dynamos that operates without the need of a net kinetic helicity.
Instead, a background shear flow coupled with nonhelical turbulence, is deemed sufficient \citep{Brandenburg2005, Yousef2008}.
A leading theoretical framework to explain this phenomenon is the shear-current effect \citep{RogachevskiiKleeorin2003, RogachevskiiKleeorin2004, RaedlerStepanov2006, RuedigerKitchatinov2006, Pipin2008, SridharSubramanian2009, SridharSingh2010, SinghSridhar2011, SquireBhattacharjee2015pre, ZhouBlackman2021, Skoutnev+2022}, which postulates a dynamo driver rooted in an anisotropic turbulent magnetic diffusivity tensor.

While the shear-current effect has been theorized to exist, its numerical substantiation has been a subject of debate \citep{Brandenburg2008, SquireBhattacharjee2015prl, SquireBhattacharjee2015apj, SquireBhattacharjee2016, Kapyla2020, Kapyla2021}. An alternative theoretical approach to addressing the shear dynamo problem is the incoherent $\alpha$ effect, wherein the interplay of shear and a fluctuating $\alpha$ coefficient with a zero mean is considered.
The concept of incoherent $\alpha$ entering the mean-field induction equation dates back to the seminal works by \cite{Kraichnan1976} and \cite{Moffatt1978},
who demonstrated that $\alpha$ fluctuations can effectively act as a negative turbulent diffusivity and contribute as an effective drift velocity if they are statistically inhomogeneous, especially when $\alpha$ fluctuations are delta-correlated.
For dynamo action to occur, it is crucial that the fluctuations are strong and coherent enough, satisfying $\tau_\alpha \alpha_\text{rms}^2\gtrsim\beta$, where $\tau_\alpha$ is the correlation time and $\alpha_\text{rms}$ is the amplitude of $\alpha$ fluctuations, and $\beta$ represents the turbulent diffusivity \citep{Kraichnan1976}.

In a related study \cite{Jingade2018}, the authors extended the work of \cite{Kraichnan1976} to include non-zero correlation times for $\alpha$ fluctuations in the presence of a background shear flow. The study showed that dynamo action is possible when there are large-scale spatial inhomogeneities in the amplitudes of the $\alpha$ fluctuations. The validity of this result in the doubly-averaged mean field depends on the time-scale separation between velocity fluctuations and $\alpha$ fluctuations, and we aim to explore and elaborate on this aspect in the present work.

In light of the limitations of \cite{Jingade2018} which is confined to small correlation times, the work of \cite{Jingade2018} was further extended to arbitrary correlation times for the velocity field using renovating flows  \citep{GB92}, taking into account the background shear self-consistently.
\cite{JingadeSingh2021} explored helicity fluctuations of the random velocity fields, also considering the idea that helicity can be correlated over times much larger than the correlation time of the velocity field, following the propositions of \cite{Kraichnan1976} and \cite{Sok97}. The main objective of this extended study was to demonstrate that large-scale dynamo action is possible when there is such time scale separation between helicity fluctuations and velocity fluctuations\footnote{%
In this work, by ``scale separation'' we always refer to the case where the correlation time of the helicity fluctuations is much longer than that of the velocity fluctuations, although the opposite case exhibits separated time scales as well,
since only the former is relevant to mean-field dynamo theories.}, in the absence of negative diffusion.
This emphasizes the significance of scale separation, a factor often overlooked in papers on mean-field electrodynamics with $\alpha$ (helicity) fluctuations and scarcely measured in numerical simulations.
Additionally, through the application of the renovating flow model in a shear flow, \cite{JingadeSingh2021} identified a critical condition for mean-field dynamo action, stipulating that $\tau_h/\tau_u>3$, where $\tau_u$ represents the correlation time of the turbulent velocity field, and $\tau_h$ represents correlation time of helicity fluctuations.
It is crucial to highlight that this conclusion was drawn from considering a single-scale flow model.
In this study, we go beyond the single-scale model and aim to explore whether large-scale flow phenomena such as shear or rotation induce a time scale separation between velocity and helicity fluctuations in direct numerical simulations of hydrodynamical turbulence.
  
The possibility of a shear dynamo driven by a stochastic $\alpha$ has also been demonstrated by directly solving for the large-scale magnetic field or its second-order moment, without explicitly deriving the contribution of $\alpha_\text{rms}$ to the turbulent electromotive force \citep{VishniacBrandenburg1997,Fedotov2006,Proctor2007,Heinemann2011,MitraBrandenburg2012,RichardsonProctor2012}. The impact of an additional stochastic component on the conventional $\alpha$ effect has been explored by \cite{NewtonKim2012} in the context of solar dynamo. Moreover, \cite{SurSubramanian2009} suggested that a fluctuating $\alpha$ might help mitigate the catastrophic quenching in the absence of a helicity flux.

In the aforementioned studies, $\alpha$ fluctuations are often prescribed (e.g., as white noise with a given amplitude or as finite correlated noise), but the origin of such fluctuations is not explicitly demonstrated. One potential source of these fluctuations arises from the violation of the Reynolds averaging rules \citep{Hoyng1987}. Alternatively, \cite{Kraichnan1976}, \cite{Moffatt1978}, and \cite{KleeorinRogachevskii2008} attribute them to the intrinsic fluctuations of the turbulent flow.

The task of determining turbulent transport coefficients in simulations is significantly facilitated by the test-field method \citep{Brandenburg2008, RheinhardtBrandenburg2010, Kapyla2020, Kapyla2021}. \cite{Brandenburg2008} observed that in shearing turbulence, all components of the $\alpha$ tensor undergo random fluctuations, following a Gaussian distribution. The amplitude $\alpha_\text{rms}$, normalized by $\beta$ and the forcing wave number $\kf$, increases with an increasing magnetic Reynolds number $\ReM$ and saturates at $\ReM=10$. However, it shows a relatively weak dependence on the shear rate at Reynolds and magnetic Reynolds numbers $\lesssim\mathcal{O}(10)$. It's worth noting that the correlation time of $\alpha$ fluctuations was not addressed in their study, and this is precisely the focus of the present work.

In this study, we aim to ascertain the correlation times of velocity and kinetic helicity directly from their time series in numerical simulations,
and determine whether the correlation time of helicity fluctuations can exceed that of the velocity fluctuations in various cases.
\revise{}{We focus on the kinematic phase on the shear dynamo problem, i.e., when the Lorentz back-reaction is negligible in the Navier-Stokes equation.}
The subsequent sections are organized as follows: Section~\ref{sec:general} briefly introduces the mean-field dynamo formalism and the $\alpha$ fluctuations.
In Section~\ref{sec:def}, we provide essential definitions of two-point correlators and describe the simulation setups. Sections~\ref{sec:rotation} and \ref{sec:shear} delve into the investigation of turbulent velocity and helicity correlation times under rotating and shearing background flows, respectively.
Section~\ref{sec:compare} discusses the implications of our findings and relates them to previous shear dynamo simulations.
Finally, we present a summary of our findings in Section~\ref{sec:conclusion}.

\section{Background}
\label{sec:general}
\subsection{Mean-field setting: $\alpha-$fluctuations }
The evolution of the magnetic field in magnetohydrodynamics is described by the induction equation, which, in the non-relativistic limit, can be expressed as
\begin{equation}
\frac{\partial \bm B}{\partial t} = \nabla \times (\bm U \times \bm B) + \eta \nabla^2 \bm B,
\label{eqn:induction_eqn}
\end{equation}
where $\bm U$ is the velocity field, $\bm B$ is the magnetic field, and $\eta$ represents the microscopic diffusivity.
Dynamo theory postulates that the action of turbulence on the seed magnetic field exponentially amplifies the magnetic field countering the microscopic diffusivity.
Furthermore in a mean-field theory, it is assumed that the typical length scale of large-scale fields is $L$ and that of the small-scale turbulent fields is $\ell_0$, with $L\gg\ell_0$.
By averaging over $\ell_0$, and assuming that the Reynolds rules of averaging hold, we can obtain the mean-field induction equation \citep{SKR1966,Moffatt1978},
\beq
\partial_t\abra{\bm B}=\del\times\left(\abra{\bm U}\times\abra{\bm B}+\bm\emf\right)+\eta\nabla^2\abra{\bmB},
\label{eqn:mf_eqn}
\eeq
where $\bm\emf=\abra{\bmu\times\bm b}$ is the turbulent electromotive force (EMF), and $\bm u$ and $\bm b$ are the fluctuating velocity and magnetic fields, respectively.

The calculation of the turbulent EMF involves solving for the fluctuating magnetic field, which in general is a complex task.
\revise{In the quasi-linear approximation, also referred to as First Order Smoothing Approximation (FOSA)}{When $\ReM<1$ or when Strohaul number ${\rm St}<1$, the quasi-linear approximation (also referred to as First Order Smoothing Approximation) is valid, and} the fluctuating field is assumed to be generated by the action of the turbulent velocity field on the initial seed mean magnetic field. With the further assumption of homogeneity and isotropy of the turbulence \citep[see, e.g.,][]{BrandenburgSubramanian2005}, an ansatz for the EMF can be obtained as follows:
\beq
\emf_i=\alpha \abra{B}_i-\beta \epsilon_{ijk}\partial_j\abra{B}_k,
\eeq
where $\alpha=-\tau_u\abra{\bm u\cdot\del\times\bm u}/3$ is related to the mean kinetic helicity $H=\abra{\bm u\cdot\del\times\bm u}$, and $\beta=\tau_u\abra{u^2}/3$ is the turbulent diffusion in the kinematic regime.
Here $\tau_u$ is the correlation time of the velocity field.
\revise{}{At $\ReM\gg1$ as relevant for astrophysical flows, other closure models like the eddy-damped quasi-normal Markovian approximation and the $\tau$-closure give similar results in the kinematic regime \citep{Pouquet+1976,BlackmanField2002}.}

As turbulence is stochastic in nature, the kinetic helicity of the flow also fluctuates in time. To propose a new dynamo mechanism in nonhelical turbulence (i.e. when $\alpha$ vanishes on average), \cite{Kraichnan1976} suggested that helicity fluctuations may have a length scale of variation, $\ell_\alpha$, larger than the turbulent scale $\ell_0$,
but still smaller than the mean-field scale $L$.
The final mean-field equation can be obtained by averaging the mean-field Equation~(\ref{eqn:mf_eqn}) again over the scale $\ell_\alpha$, i.e. a double-averaging scheme, which yields
\beq
\partial_t\abra{\overline{\bm B}} = (\beta - \alpha^2_{\rm rms}\tau_\alpha)\nabla^2\abra{\overline{\bm B}},
\label{eqn:kraichnan76}
\eeq
where $\overline{\bm\cendot}$ indicates the second averaging and $\tau_\alpha$ is the correlation time of the $\alpha$ fluctuations\footnote{In \cite{Kraichnan1976}, the terms helicity and $\alpha$ fluctuations are used interchangeably without distinction between them.}, and $\alpha^2_{\rm rms}= \overline{\alpha^2}$ is the average of the square of the $\alpha-$fluctuations. We can define the dynamo number $D_{\!\alpha} = \alpha^2_{\rm rms}\tau_\alpha/\beta$. When $D_{\!\alpha}>1$, i.e., when there is a negative diffusion in the turbulence,
the double averaged mean magnetic field grows.
$D_{\!\alpha}$ can exceed unity either when the fluctuations strength are stronger or due to the longer correlation time, and hence both these aspects are crucial for determining the dynamo action. However, in the present study, we majorly focus on the duration of the correlation time of $\alpha$-fluctuations as it is essential for the validity of the double averaged mean-field equation in \Eq{eqn:kraichnan76}.
In the shearing-turbulence simulation carried out by \cite{Brandenburg2008}, the total turbulent diffusion coefficient turns out to be positive.

Several studies have investigated the role of zero mean $\alpha$ fluctuations and shear in explaining the dynamo observed in nonhelical shear turbulence, including works by \cite{Heinemann2011}, \cite{MitraBrandenburg2012}, and \cite{Jingade2018}. In all of these models, the success of the approach depended on the existence of the temporal and spatial scale separation between the perceived fluctuations of velocity and $\alpha$, i.e.
\beq
\tau_\alpha\gg \tau_u,\ \text{and}\ \ell_\alpha\gg\ell_0.
\eeq
In an alternate approach, \cite{JingadeSingh2021} utilized the renovating flow model \citep{Dittrich1984,Sok97} to directly solve for Equation~(\ref{eqn:induction_eqn}) without explicitly determining the form of the turbulent EMF. The next subsection provides a brief introduction to this approach,
and also motivates the need of studying correlation times of individual Fourier modes.

\subsection{Renovating flow model: Helicity fluctuations}
In the renewing flows, time is divided into intervals $I_n= [n\tau,(n+1)\tau]$ of length $\tau$ for $n=0,1,2,\cdots$. A random flow
$\bfu(t,\bfx)$ is generated by choosing 
\beq
\bfu(t,\bfx) = \bfu_\tau(t-n\tau,\bfx)
\eeq
for each interval $I_n$, where $\bfu_\tau$ is chosen randomly from the ensemble $\varSigma_\tau$ of smooth flows, given by
\beq
\varSigma_\tau = \{\bfu_\tau(t,\bfx): 0\leq t< \tau\}. 
\eeq 
The velocity fields in each interval are considered to be random and statistically independent realizations of the underlying probability distribution function. The random velocity field of the ensemble $\varSigma_\tau$ is chosen to be a single-scale finite time solution of Navier-Stokes equation without the Lorentz
force\citep[see][for details]{JingadeSingh2021}.
The model velocity field $\bfu(t,\bfx;S,\bfq,h)$ hence depends on a number of parameters:
$S$ is a constant shear rate so that the background shear flow is of the form $\bmU^\text{shear}=Sx{\bm e}_y$,
$\bfq = (q_1,q_2,q_3)$ is the wave vector of $\bfu_\tau$ at initial time,
and the value of $h$ controls the relative helicity of the flow $\bfu$ which varies in the range $[-1,1]$.

Under the assumption of statistical independence of the velocity field across the intervals, the evolution of the magnetic field equation can be written separately for any given interval $[(n-1)\tau,n\tau]$ in Fourier space. Specifically, the magnetic field at time $n\tau$ can be written in terms of the initial magnetic field at time $(n-1)\tau$, after averaging over the initial randomness of the magnetic field and the statistical ensemble of the velocity field, as \citep[see][for details]{JingadeSingh2020}
\beq
\bfB(n\tau,\bfk) = {\bf G}(\tau,\bfk)\bfB_0((n-1)\tau,\bfk),
\label{eqn:RF}
\eeq
 where $\bf G$ is a response tensor that describes the average deformation of fluid particle trajectories over the interval $\tau$ due to the velocity field statistics. \revise{}{The response tensor is obtained by neglecting the diffusion term in the induction equation. As a result, the magnetic field at specific realizations exhibits fine structures similar to those at large $\ReM$. However, the averaging over the velocity statistics introduces the effective turbulent diffusivity for the growing magnetic field, smoothing the field over the turbulent velocity scale.}

In the renovating flow model presented by \cite{JingadeSingh2021}, the idea of helicity fluctuations introduced by \cite{Kraichnan1976} and \cite{Sok97} is incorporated. In this model, the helicity of the flow is assumed to take independent and random values in successive time intervals, while the durations of these intervals are fixed. It is also assumed that the renovation time of helicity, $\tau_h$, is an integral multiple of the renovation time of the velocity field, $\tau$, i.e., $\tau_h= m\tau$, where $m \geq 2$ and $\tau$ is the correlation time of the flow. In \Eq{eqn:RF}, the response tensor is averaged over the statistics of the velocity field, except for the helicity parameter $h$. To incorporate the helicity fluctuations and its time scale separation in the averaging, the response tensor from adjacent intervals are multiplied together while the helicity is fixed over the $m$ intervals. Later, the product of the response tensor is averaged over the helicity statistics to obtain the double-averaged mean field. Thus, the magnetic field at time $n\tau$ can be expressed in terms of the magnetic field at $(n-m)\tau$ as
\begin{align}
 \bfB(n\tau,\bfk) = & {\bf {\cal G}}(\tau,\bfk)\bfB_0((n-m)\tau,\bfk) \\
 {\bf {\cal G}}(\tau,\bfk) = & \left\langle {\bf G}(\tau,\bfk_{n-1}).....{\bf G}(\tau,\bfk_{n-m}) \right\rangle_h\, ,
\end{align}
where ${\bf {\cal G}}(\tau,\bfk)$ is the double averaged response tensor, whose eigenvalues determine the growth rate and cycle period of the dynamo wave \citep[see][for details]{JingadeSingh2021}. It is shown a nonhelical large-scale dynamo can be generated only when the ratio between the time scales is $\geq3$.  

\subsection{Sources of helicity fluctuations}
\label{sec:source_of_h}

In non-helical turbulence, the kinetic helicity fluctuates around zero, and these fluctuations have two primary contributions. The first contribution arises from the non-equivalence of volume averaging and ensemble averaging. In renovating flows, averaging is conducted over the ensemble of the velocity field, strictly adhering to Reynolds' averaging rule. Conversely, in mean-field theory, averaging is performed over a length scale or a finite volume, leading to an approximate adherence to Reynolds' rules, as argued by \cite{Hoyng1987, Hoyng1988} and \cite{ZhouBlackmanChamandy2018}.
In the work by Hoyng, the resulting error terms manifest as a forcing term in the mean-field equation, giving rise to fluctuations in the transport coefficients, whereas in the work by \cite{ZhouBlackmanChamandy2018}, the deviation from Reynolds' rules introduces additional terms in the mean-field equations, but these appear as spatial gradients of the mean fields.
Since volume average is not strictly equivalent to an ensemble average, the volume-averaged mean helicity $H$ fluctuates over time in simulations of non-helical turbulence, although typically with relatively small amplitudes \revise{}{(see Figure~1 in the supplementary material for the evolution of the volume-averaged helicity)}.
We denote the correlation time of $H$ as $\tau_H$, which serves as our first analogue to the correlation time of $\alpha$ fluctuations.

\revise{}{Note that the volume-averaged helicity scales as $V^{-1/2}$, where $V$ is the volume of the box, assuming the helicity fluctuations within each eddy remain consistent regardless of the volume. Furthermore, if the helicity fluctuations in each eddy are independent and identically distributed with a uniform correlation time $\tau_0$, then the correlation time of the mean helicity will also be $\tau_0$, independent of the volume. In this study, we examine how the correlation time varies with rotation or shear rate, keeping the volume fixed. The implications of finite volume, such as a scale-dependent dynamo growth rate, could be significant for astrophysical dynamos and will be explored in future work.}

The second contribution to helicity fluctuations arises from the intrinsic randomness of turbulence, where the helicity associated with each Fourier mode fluctuates around zero due to the nonlinear interaction of waves. Consequently, our second analogue to the correlation time of helicity fluctuations pertains to individual flow modes rather than volume-averaged quantities.
Compared to $\tau_H$, this idea aligns more closely with the renovating-flow model by \cite{JingadeSingh2021}, where a single-scale flow is used to investigate the dynamo phenomenon, associating fluctuating helicity at a particular mode $\bfq$ with some correlation time.  

However, real turbulence involves multiple scales. The single-scale model of \cite{JingadeSingh2021} can be extended to multi-scale flows by constructing the velocity field from independent non-interacting helical waves, with amplitudes following specific power laws across scales \citep[see][for details]{SS17}. In turbulence's steady state, the nonlinear term only cascades energy across scales, leaving the dynamics of individual modes unaffected. This extension justifies our scale-by-scale approach to verifying dynamo action criteria.  Similar methods have been employed by \cite{LouCarloShu007} and others to study dynamo problems.  Consequently, our objective is to determine the correlation times of both velocity and helicity fields across a spectrum of wave vectors $\bm k$ and verify the dynamo criteria scale-by-scale. 

The correlation time of the helicity field is obtained by considering the helicity density in Fourier space, defined as
\beq
\tilde g(t,\bmk)=\tilde{\bmu}^*(t,\bmk)\cdot i\bmk\times\tilde{\bmu}(t,\bmk),
\eeq
where $\tilde{\bmu}$ is the Fourier transform of the velocity field. It is important to note that $\tilde g$ is a real quantity and represents the helicity of the velocity field of a particular mode $\bfk$ and at time $t$. We must distinguish $\tilde g$ from the Fourier transform of the helicity density in configuration space, $h=\bmu\cdot\del\times\bmu$, which is \textit{not} used in our analysis.

\section{General definitions and numerical methods}
\label{sec:def}
\subsection{Velocity and helicity auto-correlations}
\label{sec:VH_autocor}

In theoretical analysis, ensemble averages are often used to define averages of a quantity in turbulence. However, in simulations, ensembles are often not available, and the turbulent field at different times can be treated as copies of the ensemble, provided the turbulence is stationary. In our simulations, we perform the statistical averaging of the velocity field over the time. The equivalence between the ensemble average discussed in the previous section and the temporal average is ensured by the ergodic theorem. The correlation times of velocity and helicity fluctuations are computed through their auto-correlation functions. We define the auto-correlation function of a velocity field $\bfu(t,\bmx)$ at the space point location $\bfx$ as
\beq
{\Cfull}_{u}(\tau,\bmx)\equiv\int_{t_0}^\infty\text{d}t\ 
\left[\bfu(t+\tau,\bmx)-\bar{\bfu}(\bmx)\right]{\bm\cdot}
\left[\bfu(t,\bmx)-\bar{\bfu}(\bmx)\right],
\label{eqn:Corr_f_x}
\eeq
where $t_0$ represents the initial time of the simulations when turbulence has become stationary and overbar denotes a time average.
Similarly, the auto-correlation of the Fourier transform of the velocity is defined as
\beq
{\Cfull}_{\tilde{u}}(\tau,\bmk)\equiv\int_{t_0}^\infty\text{d}t\ 
\left[\widetilde{\bfu}(t+\tau,\bmk)-\bar{\widetilde{\bfu}}(\bmk)\right]{\bm\cdot}
\left[\widetilde{\bfu}(t,\bmk)-\bar{\widetilde{\bfu}}(\bmk)\right]^*,
\label{eqn:Corr_f_k}
\eeq
where the asterisk indicates the complex conjugate.

We also define the auto-correlations of the mean helicity as
\beq
{\Cfull}_{H}(\tau) = \int_{t_0}^\infty\text{d}t\ 
\left[H(t+\tau)-\bar{H}\right]{\bm\cdot}
\left[H(t)-\bar{H}\right],
\label{eqn:CH0D}
\eeq
where $H(t) = \int \ud^3 x\ \bfu(t,\bfx)\cendot {\bm\omega}(t,\bfx) = \int \ud^3 k\ \tilde{g}(t,\bfk) $, and $\bm\omega=\del\times\bmu$ is the vorticity field.
Here $H$ has a vanishing ensemble-average mean,
but in simulations they fluctuate around zero as discussed in Section~\ref{sec:source_of_h}.
The auto-correlation of the helicity density $\widetilde{g}(\bfk, t)$ in the Fourier space  is
\beq
{\Cfull}_{\tilde{g}}(\tau,\bmk)\equiv\int_{t_0}^\infty\text{d}t\ 
\left[\widetilde{g}(t+\tau,\bmk)-\bar{\widetilde{g}}(\bmk)\right]{\bm\cdot}
\left[\widetilde{g}(t,\bmk)-\bar{\widetilde{g}}(\bmk)\right]^*.
\label{eqn:Corr_g}
\eeq

Finally, we define the volume-integrated correlation function as
\beq
\Cu{}(t)=\int \text{d}^3x\ {\Cfull}_{u}(t,\bmx) = \int \frac{\text{d}^3k}{(2\pi)^3}\ {\Cfull}_{\tilde{u}}(t,\bfk)\,. 
\label{eqn:Cu0D}
\eeq
The second equality is due to Parseval's theorem. 

\begin{figure}
\centering
\includegraphics[width=\columnwidth]{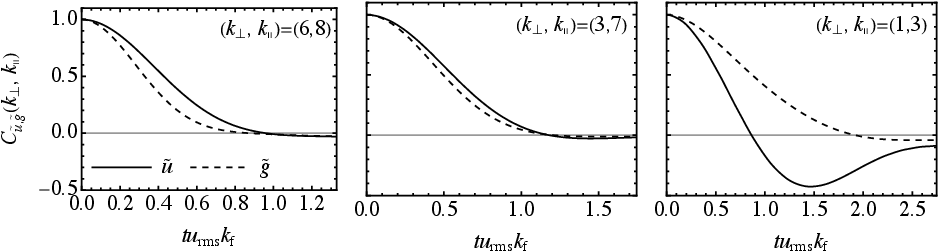}
\caption{Three typical auto-correlation curves for run R3, demonstrating how the correlation times of the velocity and helicity fluctuations vary with the wave numbers parallel ($k_\parallel$) and perpendicular ($k_\perp$) to the rotation axis.}
\label{fig:R3_3ac}
\end{figure}

\Fig{fig:R3_3ac} displays three typical auto-correlation curves of run R3  (further details can be found in Section~\ref{sec:rotation}). The correlation time of an auto-correlation curve is determined by fitting its initial positive part with an exponentially decaying curve and identifying the time at which its value reaches $1/e$ of its maximum. We denote the correlation time of ${\Cfull}_I$ by $\tau_I$, where $I\in\{u,\tilde{u}, g, \tilde{g}, H\}$.

\subsection{Distinguishing different helicity time scales}
\label{sec:2tauH}
A volume-integrated correlation function for the helicity modes is defined by integrating the helicity auto-correlation function over the wave vector space, similar to Equation~(\ref{eqn:Cu0D}),
\beq
\Cg{}(t)=\Ctg{}(t)=\int\frac{\text{d}^3k}{(2\pi)^3}\ {\Cfull}_{\tilde{g}}(t,\bmk).
\label{eqn:Ctg0D}
\eeq
It should be noted that although one might expect the correlation times of $\Cg{}$ and $C_H$ to be similar, this is not necessarily the case.
To see the reason, we assume that $\tilde g(\bmk)$ has a vanishing mean for the moment and note that
\begin{align}
&C_H=\int\text{d}t'\int\frac{\text{d}^3k_1}{(2\pi)^3}\ 
\tilde g(t',\bm k_1)
\int\frac{\text{d}^3k_2}{(2\pi)^3}\ \tilde g(t'+t,\bm k_2),
\label{eqn:CH0D_kf}\\
&\Cg{}=\int\frac{\text{d}^3k}{(2\pi)^3}\int \text{d}t'\ 
\tilde g(t',\bm k)\tilde g(t'+t,\bm k).
\label{eqn:tautg0D_kf}
\end{align}
It is clear from Equations~(\ref{eqn:CH0D_kf}) and (\ref{eqn:tautg0D_kf}) that $C_H$ is the auto-correlation of the sum of all modes, whereas $\Cg{}$ is proportional to the mean of the auto-correlations over all modes.

Both $\taug$ and $\tau_H$ have been utilized in dynamo models, each with its own significance.
In the original negative diffusion model proposed by \cite{Kraichnan1976}, the time scale for fluctuations of the $\alpha$ coefficient was closely related to $\tau_H$. Similarly, stochastic $\alpha$ dynamo models, such as those by \cite{VishniacBrandenburg1997} and \cite{RichardsonProctor2012}, also adopt a time scale associated with $\tau_H$.
On the other hand, researchers studying turbulence using random waves, as exemplified by  \cite{JingadeSingh2021},  focus on the damping time of a single helical wave and therefore utilize $\taug$. In our study, we present results for both $\taug$ and $\tau_H$, recognizing the importance of considering both perspectives.

\subsection{Numerical setups and methods}
\label{sec:numerics}
\revise{}{Focusing on the kinematic dynamo regime and neglecting Lorentz forces,}
we perform compressible isothermal hydrodynamic simulations of isotropically and nonhelically forced turbulence using the  publicly available {\sc Pencil Code} \citep{pencilcode2021}.
The equations to be solved are
\begin{align}
&\text{D}_t\ln\rho=-\del\cdot\bmu,
\label{eqn:dlnrho_dt}\\
&\text{D}_t\bmu=Su_x{\bm e}_y-c_\text{S}^2\del\ln\rho-2\bm\Omega\times\bmu
+\frac{1}{\rho}\del\cdot\left(2\rho\nu\bm S\right)+\bm f,
\label{eqn:du_dt}
\end{align}
where $\text{D}_t=\partial_t+\bmu\cdot\del+Sx\partial_y$,
$S$ is the shear rate of the shear flow $\bm U^\text{shear}=Sx{\bm e}_y$,
$c_\text{S}$ is a constant sound speed, $\nu$ is the viscosity, and $S_{ij}=(\partial_i u_j+\partial_j u_i)/2-\delta_{ij}\del\cdot\bmu/3$ is the rate-of-strain tensor.
In the burgulence cases, the density equation~(\ref{eqn:dlnrho_dt}) is dropped and the second term on the right-hand side of Equation~(\ref{eqn:du_dt}) is set identically to zero, turning the Navier-Stokes equations into Burgers' equation.
The rotation rate $\bm{\Omega} = \Omega \bm{\ez}$ is applied along the $z$-direction, and the random force $\bm{f}$ takes the form of a nonhelical plane wave \citep[see][for details]{Axel2001}.
The forcing wave vector has a fixed magnitude $\kf$, but its phases and directions change at each time step.
The simulation box has dimensions $L^3 = (2\pi)^3$ with periodic boundary conditions in all the directions or shear-periodic (in case of shear) boundary conditions in the $x$-direction.
In all runs, the forcing wave number is $\kf = 5$, and the Mach number is approximately $0.1$.
A summary of the runs is provided in Table~\ref{tab:runs}, which includes the Reynolds number $\ReN = \urms/(\nu\kf)$, the Coriolis number $\Co = 2\Omega/(\urms\kf)$, and the dimensionless shear number $\Sh=-S/(\urms\kf)$.
For the rotating turbulence cases R1 to R5, we compute the Coriolis number by excluding the $k<\kf$ modes when computing the root-mean-square velocity $\urms$ (see Section~\ref{sec:rotation}).
We denote the Coriolis number computed using this definition of $\urms$ as $\Cokf$.
For runs~R3 to R5, their Coriolis numbers are sufficiently large to allow for condensation of large-scale vortices, and therefore their Reynolds numbers are apparently higher than those of runs~R1 to R2.
At even higher values of $\Cokf$, the flow becomes quasi two-dimensional and dynamos will not be supported \citep{Cowling1933}. Therefore, we exclude the exploration of high Coriolis numbers.

In this study, particular attention is given to ensuring the hydrodynamic stability of the flow in all simulations.
The choice of shear rate and rotation rate is made such that the flow remained stable throughout the simulations. This criterion for stability, $-\infty < {-S}/{\Omega} < 2$, as discussed in \cite{Salhi2002} and \cite{Balbus2006}, has been taken into consideration.
The presence of instabilities in the flow could lead to vorticity dynamos \citep{Kapyla2009}, i.e., the generation of additional mean flows.
Such mean flows would render the system non-ergodic.
To ensure the validity of measurements that require time-stationarity such as the temporal average of the correlation functions, it is essential to prevent such instabilities from affecting the measurements.

\begin{table}
\begin{center}
\def~{\hphantom{0}}
\begin{tabular}{lccccccc}
\hline
Run & $N^3$ & $\Ma$ & $q=-S/\Omega$ & $\ReN$ & $\Co$ or $\Cokf$ & $\Sh$ & $\kze$\\
\hline
NH & $256^3$ & $0.12$ & $/$ & $45.93$ & $0$ & $0$ & /\\
\hline
R1 & $256^3$ & $0.02$ & $0$ & $32.17$ & $0.43$ & $0$ & 0.35\\
R2 & $256^3$ & $0.03$ & $0$ & $33.14$ & $0.86$ & $0$ & 5.19\\
R3 & $256^3$ & $0.07$ & $0$ & $90.34$ & $2.03$ & $0$ & 14.24\\
R4 & $256^3$ & $0.08$ & $0$ & $108.06$ & $3.20$ & $0$ & 25.68\\
R5 & $256^3$ & $0.09$ & $0$ & $115.94$ & $4.29$ & $0$ & 40.75\\
\hline
S1 & $256^3$ & $0.12$ & $-4$ & $46.86$ & $0.03$ & $0.07$ & /\\
S2 & $256^3$ & $0.14$ & $-4$ & $55.34$ & $0.11$ & $0.22$ & /\\
S3 & $256^3$ & $0.16$ & $-4$ & $64.62$ & $0.29$ & $0.57$ & /\\
S4 & $256^3$ & $0.13$ & $-4$ & $50.47$ & $0.76$ & $1.52$ & /\\
S5 & $256^3$ & $0.11$ & $-8$ & $42.15$ & $0.91$ & $3.65$ & /\\
\hline
SB1 & $256^3$ & / & $-4$ & $15.10$ & $0.04$ & $0.08$ & /\\
SB2 & $256^3$ & / & $-4$ & $15.31$ & $0.15$ & $0.31$ & /\\
SB3 & $256^3$ & / & $-4$ & $14.21$ & $0.50$ & $1.00$ & /\\
SB4 & $256^3$ & / & $-4$ & $11.95$ & $1.24$ & $2.48$ & /\\
\hline
K1 & $256^3$ & $0.12$ & $3/2$ & $47.62$ & $0.13$ & $0.10$ & /\\
K2 & $256^3$ & $0.13$ & $3/2$ & $49.80$ & $0.74$ & $0.56$ & /\\
K3 & $256^3$ & $0.12$ & $3/2$ & $45.60$ & $1.69$ & $1.27$ & /\\
K4 & $256^3$ & $0.11$ & $3/2$ & $41.96$ & $3.67$ & $2.75$ & /\\
K5 & $256^3$ & $0.10$ & $3/2$ & $37.36$ & $8.24$ & $6.18$ & /\\
\hline
\end{tabular}
\caption{%
A summary of runs, where $N^3$ is the resolution, $\Ma=\urms/c_\text{S}$ is the Mach number, $S$ is the shear rate, $\Omega$ is the rotation rate, $\ReN=\urms/\nu\kf$ is the Reynolds number, $\Co=2\Omega/\urms\kf$ is the Coriolis number, $\Sh=-S/\urms\kf$ is the dimensionless shear rate, and $\kze$ is the Zeman wave number.
For rotation runs~R1 to R5, the root-mean-squared velocity used for computing the Coriolis number excludes large-scale modes with $k<\kf$, and therefore their Coriolis numbers will be denoted by $\Cokf$ rather than $\Co$.
For these runs the Zeman wave number $\kze$ is also listed.
The runs SB1 to SB4 solve the Burgers' equation and the Mach number is undefined.}
\label{tab:runs}
\end{center}
\end{table}

We use two methods to generate auto-correlation curves of the velocity and helicity fluctuations.
In the first method, we output the time series data for $\tilde\bmu(t,\bmk)$ from the simulations and calculate auto-correlation curves during post-processing.
The duration of time series for all runs spans at least $200$ times the eddy turnover time $1/(\urms\kf)$ with a sampling frequency of no less than $10\urms\kf$.
For runs with the highest shear rates (S4, S5, SB4, K4, and K5), the sampling frequency is increased to a minimum of $20\urms\kf$.
This approach allows for obtaining auto-correlations for individual Fourier modes.
As large-scale flow patterns such as shear or rotation primarily influence the larger scales of the flow,  we have limited the output to the range $-2\kf\leq k_{x,y,z}\leq2\kf$, which has been confirmed to encompass the energy-dominant modes.

The first method needs high data output frequency and therefore is only applied to a limited range of wave modes.
In the second method, correlation functions are computed on-the-fly and for all the wave modes, though this approach results in a smaller sample size for time averaging.
The method is specifically used to obtain shell-averaged correlation functions in the Fourier space.
To implement this,  we regularly update an time-independent auxiliary field $\tilde{\bm v}(\bmk)$ to match $\tilde\bmu(t,\bmk)$ at specific time intervals (every $\Delta t=50$ in code units, roughly $27$ times the eddy turnover time of the isotropic run).
The field $\tilde{\bm v}(\bmk)$ is kept fixed between updates.
At a given update time $t_0$, we set $\tilde{\bm v}_0(\bmk)$ to be the value of $\tilde{\bm u}(t_0, \bmk)$. During the interval $t_0 \leq t < t_0 + \Delta t$, the velocity auto-correlation between $\tilde{\bm v}_0(\bmk)$ and $\tilde{\bm u}(t, \bmk)$ is computed as:
\beq
{\Cfull}_{\tilde u}(t,\bmk)=
\tilde{\bm v}^*_0(\bmk){\cdot}\tilde{\bm u}(t_0+t,\bmk)=
\tilde{\bm u}^*(t_0,\bmk){\cdot}\tilde{\bm u}(t_0+t,\bmk),
\ 0\leq t\leq \Delta t.
\eeq
Similarly, the helicity correlation ${\Cfull}_{\tilde g}$ is computed. These correlation curves are then averaged at discrete intervals. 
 
While this method allows us to analyze modes at the largest wave numbers, it limits the number of averaged ensemble members to $T / \Delta t$, where $T$ is the overall time duration of the simulation.

\begin{figure}
\centering
\includegraphics[width=0.5\columnwidth]{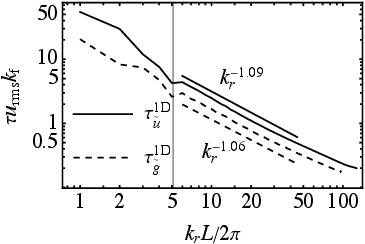}\\
\caption{For run NH, correlation times of shell-averaged auto-correlations for velocity (solid) and helicity (dashed) fields.
The vertical line indicates the forcing wavenumber at $\kf=5$.}
\label{fig:Iso_1D}
\end{figure}

In Figure~\ref{fig:Iso_1D}, we present correlation times for shell-averaged quantities in isotropic nonhelical turbulence. The shell-averaged correlation functions are defined by
\beq
C_{\tilde u,\tilde g}(t,k_r)=\frac{1}{4\pi}\int \sin\theta\,\text{d}\theta\,\text{d}\phi\ {\Cfull}_{\tilde u,\tilde g}(t,\bmk),
\label{eqn:C_1D}
\eeq
where $(k_r,\theta,\phi)$ are the spherical coordinates in Fourier space. Subsequently, the correlation times for each wave number are fitted as $\tautu{1D}(k_r)$ and $\tautg{1D}(k_r)$ for the velocity and helicity auto-correlations, respectively.
The identified $k_r^{-1}$ power law for $\tautu{1D}$ in Figure~\ref{fig:Iso_1D} suggests that the Eulerian correlation time is primarily influenced by the sweeping effect, where small-scale eddies are transported by larger-scale eddies. Consequently, $\tautu{1D}\propto 1/(\urms k_r)$, aligning with findings from prior theoretical and numerical studies \citep{ChenKraichnan1989,SanadaShanmugasundaram1992,Favier2010,Clark2015}. 

We have also computed the correlation time of volume-integrated velocity or helicity correlation, along with that of the total helicity using Equations~(\ref{eqn:CH0D}), (\ref{eqn:Cu0D}), and (\ref{eqn:Ctg0D}). The obtained values are:
\beq
\tau_u\urms\kf=3.11,\ 
\tau_H/\tau_u=2.42,\ 
\taug /\tau_u=0.61.
\label{eqn:Iso_0D}
\eeq
These values serve as a set of fiducial references, which we will compare to the results obtained when rotation or shear is introduced.
\cite{BrandenburgSubramanian2005AAp} previously reported a Strouhal number $\text{St}=\tau_u\urms\kf$ around unity for the $\bmu\cdot\bm\omega$ correlation, determined by dividing the $\alpha$ coefficient from a test-field method by one-third of the mean kinetic helicity. Our value of $3.11$ is not fundamentally at odds with this, mainly due to the different methods and, consequently, the different definitions of $\tau_u$ employed.

In the next two sections, we present the measured correlation times for velocity and helicity fluctuations in rotating and shearing turbulence, respectively.
A more detailed discussion of the implications for the shear dynamo problem along with the comparison with previous simulation results is provided in Section~\ref{sec:compare}.

\section{Rotating flows}
\label{sec:rotation}
In this section, we compare the correlation times of velocity and helicity fluctuations in rotating turbulence. Rotating turbulence is known to form a condensate through the inverse cascade of energy and, in a steady state, counter-rotating vortices are formed to cancel the effect of global rotation\footnote{This is a consequence of the generalisation of the Kelvin vorticity theorem in a rotating frame and it is known as \emph{Bjernkne's theorem}.} \citep{Kraichnan1967,Bartello1994,Alexakis2015}. The rotation tends to suppress velocity gradients along the direction of rotation, making the flow quasi-two-dimensional at sufficiently large Coriolis numbers.

\begin{figure}
\centering
\includegraphics[width=0.4\columnwidth]{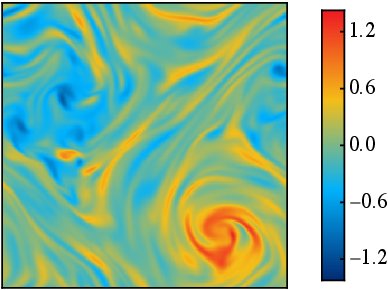}
\caption{The $z$ component of the vorticity field (normalized by $\kf\urms$) at the $z=0$ slice for run R4 in the steady state.}
\label{fig:R4_oo3_z0}
\end{figure}

In Figure~\ref{fig:R4_oo3_z0}, the $z$ component of the vorticity field at the $z=0$ slice for run R4 is plotted at the steady-state regime of the simulation, and we observe asymmetric cyclonic and anti-cyclonic vortices aligned in the direction of rotation [also seen in \cite{SeshasayananAlexakis2018}, \cite{DallasTobias2018} and the references therein].
The vortices display an asymmetry in their positive and negative strengths, with cyclonic vortices having a comparatively higher strength than anti-cyclonic vortices. This is because one of the vortices is \textit{Rayleigh} stable for a given sign of rotation and strength \citep{Tritton1985}. The formation of these vortices will alter the estimate of the correlation time of both velocity and helicity fluctuations. 

Since our interest is in the large-scale dynamos, we focus on the turbulent flow and therefore exclude the $k<\kf$ modes when computing the root-mean-square velocity $\urms$.
We denote the Coriolis number computed using this definition of $\urms$ as $\Cokf$.
At sufficiently large wave numbers where the isotropic eddy turnover rate is larger than the rotation rate,
the turbulence is expected to recover the Kolmogorov scaling with a power-law energy distribution of $k^{-5/3}$.
The scale at which this transition occurs is referred to as the Zeman scale, given by $\kze=\left(\Omega^3/\epsilon\right)^{1/2}$, where $\epsilon = \langle\bfu\cendot{\bm f} \rangle$ is the energy injection rate.
This is the scale at which the rotation rate $\Omega$ equals the eddy turnover rate $k {\tilde u}(k)$ \citep{Zeman1994}.

The properties of the flow are different in the direction of rotation ($\hat k_\parallel$) and in the direction perpendicular to it ($\hat k_\perp$), presenting axial symmetry.
We therefore obtain the correlation functions by performing the azimuthal average as 
\beq
C^{\rm axi}_{\tilde u, \tilde g}(t,k_{\perp},k_{\parallel}) = \frac{1}{2\pi}\int {\rm d}\phi\; {\Cfull}_{\tilde u,\tilde g}(t,\bmk)\, .
\eeq
Additionally, we average over modes with the same $|k_{\parallel}|$, because the correlation functions of velocity or helicity are invariant under coordinate reflection over the $k_\parallel=0$ plane. 

\begin{figure}
\centering
\includegraphics[width=\columnwidth]{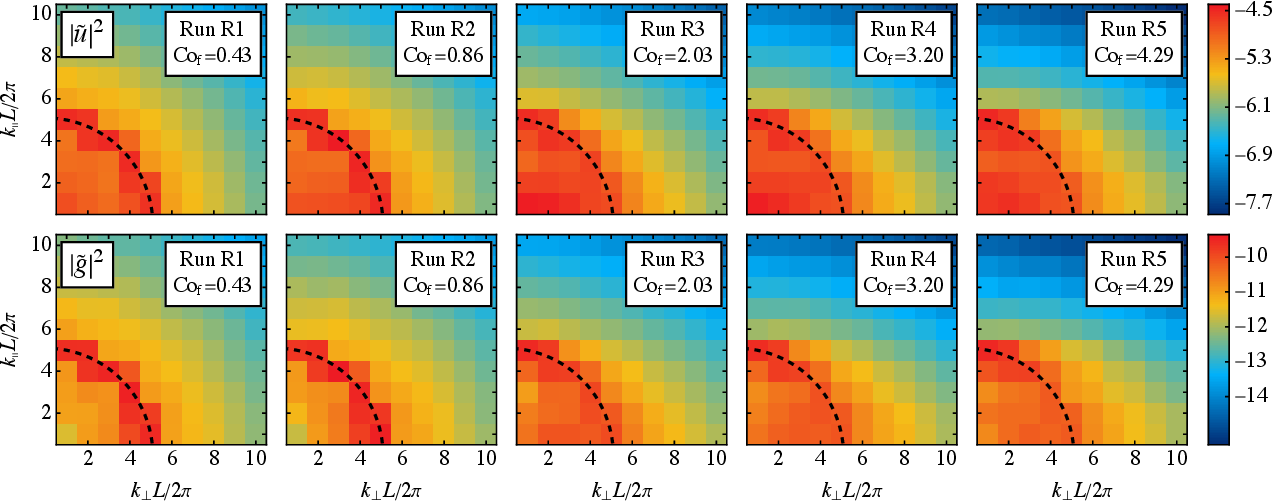}
\caption{For rotating turbulence, the velocity ($|\tilde u|^2$, top) and helicity ($|\tilde g|^2$, bottom) energy densities on a $\log_{10}$ scale.
The dashed curves indicate the forcing wave number $\kf=5$.}
\label{fig:R_density_Axi}
\end{figure}

Note that $C^{\rm axi}_{\tilde u,\tilde g}(t=0,k_\perp,k_\parallel)$ are just the velocity
and helicity energy densities (i.e. quadratic in $\tilde u$ or $\tilde g$),
which we show in Figure~\ref{fig:R_density_Axi}.
For the nearly isotropic cases R1 and R2, both velocity and helicity energy densities concentrate
near the forcing scale $\kf=5$.
At high rotation rates, the velocity energy density smears into the low-$k$ region manifesting the large-scale vortices,
and the helicity modes favor to reside along the $k_\parallel$ direction.

The correlation times, denoted as $\tautu{axi}(k_\perp,k_\parallel)$ and $\tautg{axi}(k_\perp,k_\parallel)$ for velocity and helicity fluctuations, respectively, are estimated by fitting the positive part of the correlation function to an exponential curve, as described in Section~\ref{sec:VH_autocor}.
Figure~\ref{fig:R3_3ac} displays three representative auto-correlation curves for run R3, which show that helicity fluctuations become increasingly coherent over time as we move to smaller wave number regions in the Fourier space. This coherence is a result of the formation of cyclonic and anti-cyclonic vortices in the rotating turbulence. Visualizing this phenomenon of increasing correlation time of helicity, fluid elements entering these vortices would follow a spiraling trajectory, and the helicity sign of these elements would be contingent upon the angle of entrance relative to the vortices' axis. 

\begin{figure}
\centering
\includegraphics[width=\columnwidth]{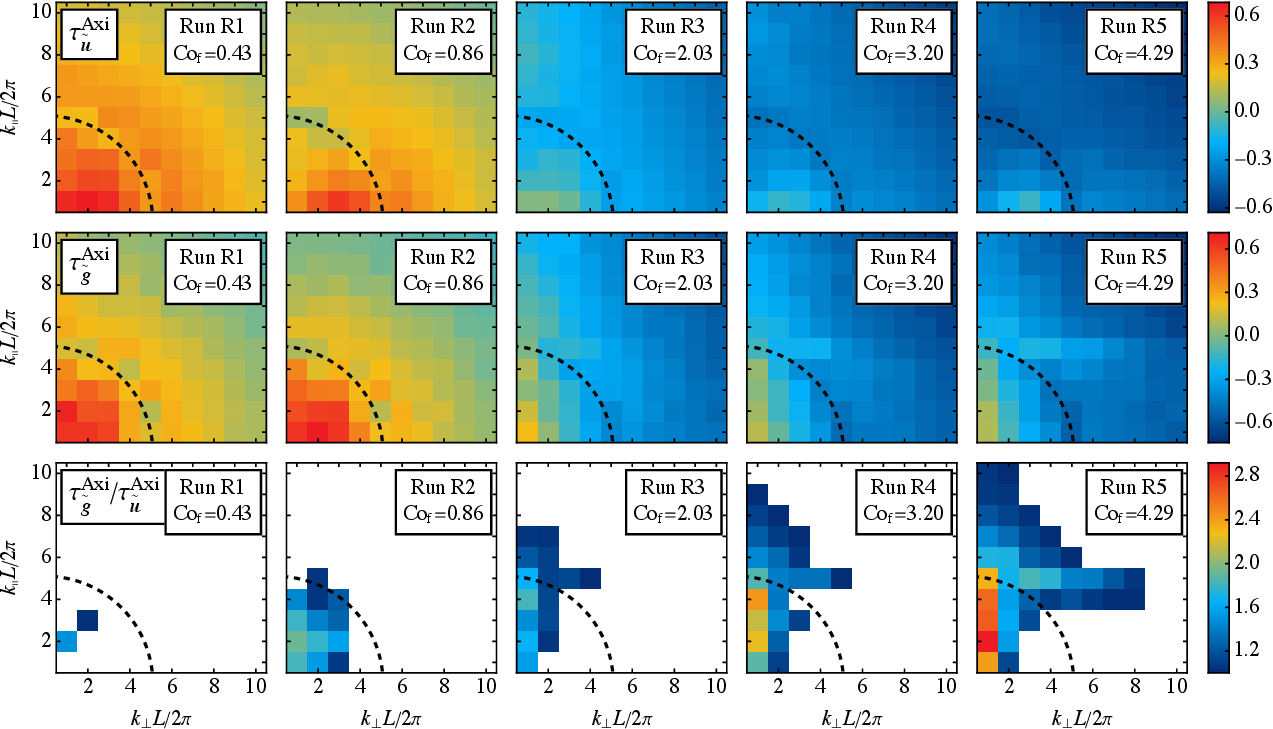}
\caption{For rotating turbulence, the first and second rows show the correlation times of velocity and helicity fluctuations in the $k_\perp-k_\parallel$ plane, respectively.
The correlation times are normalized by the eddy turnover time $1/(\kf\urms)$ and displayed on a $\log_{10}$ scale. Here $\urms$ excludes large-scale modes with wave numbers $k < \kf$ and $\kf$ denotes the forcing scale, indicated by the black dashed curve at $\kf=5$.
The third row shows the ratio of these two times on a linear scale, which is displayed only for the region where the ratio is equal to or greater than unity.}
\label{fig:R2D}
\end{figure}

The correlation times for a range of $(k_\perp,k_\parallel)$ pairs are displayed in the first two rows of Figure~\ref{fig:R2D}.
In the third row of the same figure, the ratio of the helicity to velocity coherence times, $\zt{axi}\equiv\tautg{axi}/\tautu{axi}$, is displayed for regions where $\zt{axi}\geq1$,
highlighting the modes for which the coherence time of helicity fluctuations is longer than that of velocity fluctuations, therefore potentially capable of driving large-scale dynamos \citep{JingadeSingh2021}.
This should be contrasted with Figure~\ref{fig:Iso_1D} and Equation~(\ref{eqn:Iso_0D}), where the opposite trend $\zeta<1$ is seen for all the wave numbers in the isotropic case.
With increasing rotation rate, an increase in both the area where $\zt{axi}\geq1$ and the maximum values of $\zt{axi}$ can be observed.
The Zeman scale, which is the scale at which the rotation rate becomes equal to the eddy turnover time, has also been computed for all the runs and is provided in Table~\ref{tab:runs}. As seen from the table, the scales where $\zt{axi}=\tautg{axi}/\tautu{axi}>1$ are within the Zeman scale for runs~R1 and R2, indicating that the difference in coherence times is a result of the rotation rate having a significant impact at these scales.
For runs~R3 to R5, the Zeman wave numbers are out of the plotted range.
Nevertheless, the condition $\zt{axi}>1$ predominantly occurs within the region $k<\kze$.

\begin{figure}
\centering
\includegraphics[width=0.5\columnwidth]{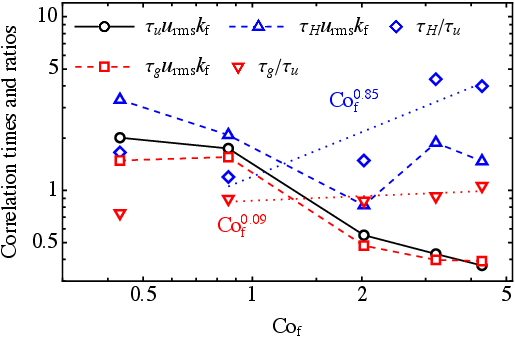}\\
\caption{For rotating turbulence, the volume-averaged
velocity and helicity correlation times, and their ratios.
Here $\Cokf$ is the Coriolis number $2\Omega/(\urms\kf)$ with $\urms$ excluding the modes with $k<\kf$.
The dotted blue and dashed red lines are the fitted power-law relations for $\tau_H/\tau_u$ and $\tau_g/\tau_u$, respectively.}
\label{fig:R0D}
\end{figure}

For volume-averaged time scales,
in order to exclude the impact from the secondary mean flows generated by large scale vortices,
we calculated the volume-integrated auto-correlation for the small-scale fields by summing over all the Fourier modes with $k\geq\kf$,
\begin{align}
&\Cu{}(t)=\int_{|\bm k|\geq\kf} \frac{\text{d}^3k}{(2\pi)^3}\ {\Cfull}_{\tilde u}(t,\bmk),\\
&C_{H}(\tau) = \int_{t_0}^\infty\text{d}t\ 
\left[H_\text{f}(t+\tau)-\bar{H_\text{f}}\right]{\bm\cdot}
\left[H_\text{f}(t)-\bar{H_\text{f}}\right],\ \text{with}\ 
H_\text{f}=\int_{|\bm k|\geq\kf} \frac{\text{d}^3k}{(2\pi)^3}\ \tilde g(\bmk),
\\
&\Cg{}(t)=\int_{|\bm k|\geq\kf} \frac{\text{d}^3k}{(2\pi)^3}\ {\Cfull}_{\tilde g}(t,\bmk).
\label{eq:sscale_cor}
\end{align}
This analysis includes all the small-scale modes and is complementary to \Fig{fig:R2D}, which only considered wave numbers up to $2k_f$. 
In \Fig{fig:R0D}, we present the correlation times obtained from these correlators.
As the Coriolis number $\Cokf$ increases, all the three dimensionless times scales, $\tau_{u,g,H}\urms\kf$ decrease, with $\tau_H\urms\kf$ decreasing at a slightly lower rate.
Meanwhile, the ratios of the time scales $\tau_{H,g}/\tau_u$ increase with increasing $\Cokf$,
though with a rather weak dependence, as indicated by the dotted lines in the figure.
The ratio $\tau_g/\tau_u$ never exceeds unity even at the largest rotation rate we have explored, whereas the ratio $\tau_H/\tau_u$ has a larger magnitude and is always above unity. The trend of the ratio $\tau_g/\tau_u$ for the modes $k>k_f$ appears to be unaffected by the rotation rate, suggesting that these modes are unimportant for large-scale dynamo action.

\begin{figure}
\centering
\includegraphics[width=\columnwidth]{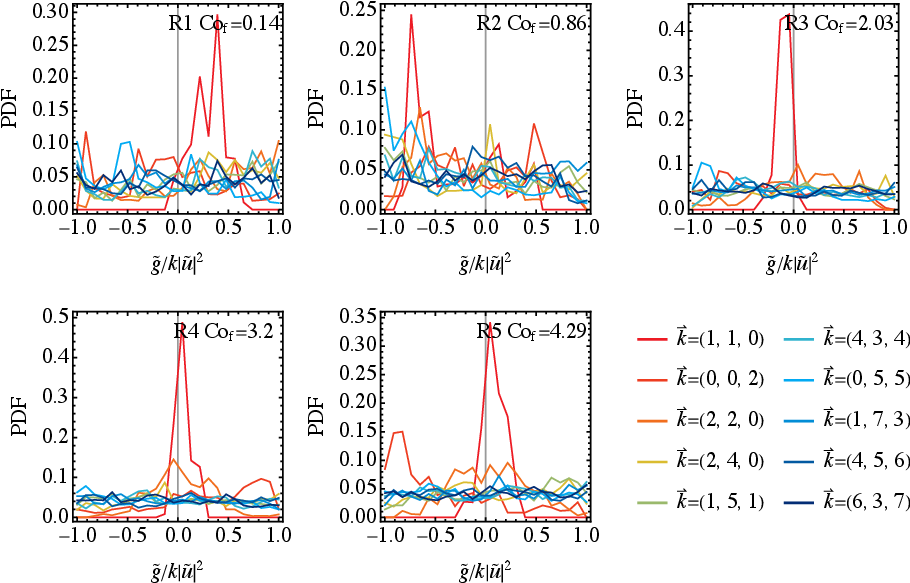}
\caption{PDFs of the normalized helicity $\tilde g/k|\tilde u|^2$ for a few modes for the rotating turbulence runs.}
\label{fig:R_pdf}
\end{figure}

In addition to the helicity correlation time, the efficiency of the incoherent $\alpha$ effect also depends on the amplitude of helicity fluctuations [see \Eq{eqn:kraichnan76}].
In \Fig{fig:R_pdf}, we depict the probability density functions (PDFs) of the relative helicity, $\tilde g/k|\tilde u|^2$, for 10 different representative modes with varying scales, across all runs.
The relative helicity is defined such that when the mode at wave number $\bm k$ is fully helical, $\tilde g/k|\tilde u|^2=\pm1$ (corresponding to positively or negatively helical cases), and it becomes $0$ when the mode is nonhelical.
Partially helical modes have relative helicities between $\pm1$.
The PDFs for the low-wave number modes ($k<\kf$) tend to have a non-zero mean, indicative of the presence of asymmetric counter-rotating large-scale vortices that affect the distribution of helicity fluctuations. On the other hand, the PDFs for the high-wave number modes ($k>\kf$) exhibit symmetry around zero and are nearly uniform in nature.

We can estimate the dynamo number, which is the ratio of the strength of the helicity fluctuations to the turbulent diffusivity [(]see the discussion below, \Eq{eqn:kraichnan76}], by utilising $\alpha_\text{rms}\simeq\tau_u H_\text{rms}/3$ (with $H_\text{rms}$ being the root-mean-square value of the fluctuating mean kinetic helicity $H$), $\beta\simeq\tau_u \urms^2/3$, and either $\tau_\alpha=\tau_H$ or $\tau_g$ as follows:
\beq
D_{H,g}\simeq\frac{\tau_{H,g}\alpha^2_\text{rms}}{\tau_u\urms^2/3}
\simeq\frac{\tau_u\tau_{H,g}\Hrms^2}{3\urms^2}.
\label{eqn:xi}
\eeq
The calculated values indicate that $D_{H,g}<2\times10^{-3}$ across all the purely rotating turbulence runs, implying that negative diffusion due to helicity fluctuations is negligible in these simulations.
For mean helicity fluctuations $H$, even though its correlation time surpasses that of velocity fluctuations, the amplitude of the helicity fluctuations are insufficient to instigate negative diffusion needed to trigger the dynamo, as predicted by \cite{Kraichnan1976}.

\section{Turbulence in shearing flows}
\label{sec:shear}
\begin{figure}
\centering
\includegraphics[width=0.4\columnwidth]{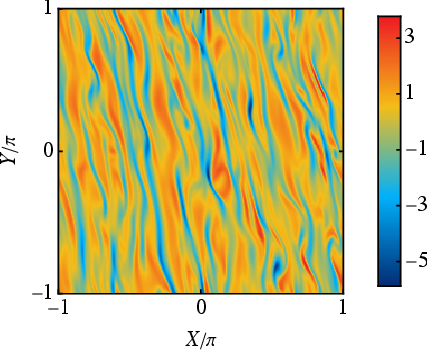}
\caption{The $z$ component of the vorticity field (normalized by $\kf\urms$) at the $z=0$ slice for run S4 at the end of the simulation.}
\label{fig:S4_oo3_z0}
\end{figure}

We now turn to the analysis of turbulence in a shearing box.
In this section, the Cartesian coordinates in the lab frame are denoted by $(\bm X,\bm Y,\bm Z)$,
and we investigate shearing flows denoted by ${\bfU}^\text{shear}=SX{\bm e}_y$ .
We also include a small yet negative rotation with $q=-S/\Omega<0$ to circumvent vorticity dynamos, as discussed in \cite{Elperin2003} and \cite{Kapyla2009}. Figure~\ref{fig:S4_oo3_z0} presents an $XY$-slice of the $Z$-component of the vorticity field, revealing the absence of significant large-scale vortices, in contrast to Figure~\ref{fig:R4_oo3_z0}. Instead, long streaky structures are evident, elongated in the direction of shear, hinting at the presence of stretched vortices aligned with the shear direction.
In this section, we will investigate how shear-induced elongation of vortices affects the velocity and helicity fluctuations.

\subsection{Shearing box and shearing frame transformation}
\begin{figure}
\centering
\includegraphics[width=0.5\columnwidth]{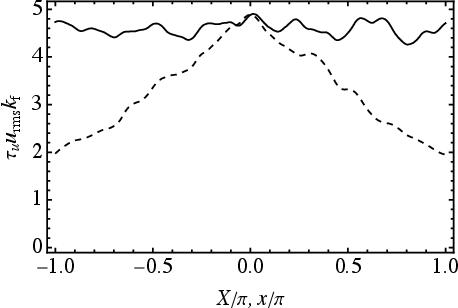}\\
\caption{For run S1,
correlation times of $YZ$-averaged velocity auto-correlations in the lab frame (dashed) and the shearing frame (solid).
Spatial homogeneity is restored in the latter case.}
\label{fig:shearing_frame}
\end{figure}

The shearing box approximation is commonly used in numerical simulations to study a small patch in a differentially rotating flow whose scale is much smaller than the local curvature of the rotating flow \citep{GoldreichLynden-Bell1965}.
When considering mean linear shear flows imposed on the turbulent velocity field, the resulting turbulence exhibits inhomogeneity and anisotropy when observed in the laboratory frame coordinates.
In particular, fluid parcels are advected by the background shear flow, making the same-point two-time correlation [Equation~(\ref{eqn:Corr_f_x})] ${\Cfull}_u(\tau,\bm X)$ have a shorter correlation time towards the boundary of the box $X=\pm\pi$;
see the dashed curve in Figure~\ref{fig:shearing_frame}.

Such decoherence effect can be removed by exploiting a fundamental symmetry of shearing flows.
An observer co-moving with the mean flow velocity $SX{\bm e_Y}$ at location $\bfX$ perceives the same turbulent velocity field as an observer in the laboratory frame at the origin. This principle, known as Galilean invariance, ensures the homogeneity of the velocity field in the shearing coordinates $(x,y,z)$ given by \citep{SridharSubramanian2009,SridharSubramanian2009rapid} 
\beq
x=X; \qquad y = Y-StX;\qquad z = Z.
\label{coor}
\eeq
The general Galilean-invariant time-stationary velocity correlation can be written in the lab frame using the symmetry property of shear flow as (see Appendix~\ref{sec:appx1}): 
\beq
\bar{\bmu^\text{lab}(t,\bm X)\cdot \bmu^\text{lab}(t',\bm X')}
={\Cfull}_{u,\text{lab}}\left(\bm X-\bm X'-S(t-t')X{\bm e_Y},t-t'\right), 
\label{eqn:Culab}
\eeq
where overhead bar denotes the time average in our case.
This correlation is time-stationary but exhibits spatial inhomogeneity along the $X$-direction.
To address this, we can transform the correlator into the shearing frame, where it becomes spatially homogeneous but loses time stationarity.
The latter can only be restored in two special slices in the Fourier space (see later for the details).
The solid curve in Figure~\ref{fig:shearing_frame} shows the velocity correlation time of the $yz$-averaged correlators, which restores both spatial homogeneity and time stationarity.

An additional advantage of using the shearing frame is the restoration of normal periodic boundary conditions, allowing for a Fourier transform in the shear-wise direction $\hat{\bm x}$. Fourier modes in the shearing frame, denoted as $(k_x,k_y,k_z)$, are equivalent to shearing-wave modes with wave vectors $\bfK=(k_x-Stk_y,k_y,k_z)$ in the lab frame. It's important to note that there would be no difference in the correlation time of the total kinetic helicity, denoted as $\tau_H$, between the two frames. This is due to the fact that $H(t)$, being a volume-averaged quantity, remains constant regardless of the frame of reference.

Without further notice, we shall always refer to shearing-frame coordinates by using $(t,x,y,z)$ in the configuration space and $(k_x,k_y,k_z)$ in the Fourier space.
The helicity density at the shearing wave vector $\bmk$ is given by $\tilde g(\bm k)=\tilde\bmu^*(\bm k)\cdot\tilde{\bm\omega}(\bm k)$, where the vorticity field $\tilde{\bm\omega} = \bfK(\bfk,t)\times\tilde{\bmu} $ is obtained by first taking the curl of $\bmu$ in the lab frame, followed by the shearing frame transformation, and then performing the Fourier transformation.

Sacrificing time stationarity in the Fourier space is unavoidable when reintroducing spatial homogeneity in the shearing frame (see Appendix~\ref{sec:appx1}). In other words, the unequal-time auto-correlation function $\abra{\tilde{\bmu}(t+t',\bmk)\cdot\tilde{\bmu}(t',\bmk)}$ not only depends on the time difference $t$ but also on $t'$. This outcome is a general consequence of Galilean invariance \citep{SridharSingh2014}. In Appendix \ref{sec:appx1}, it is demonstrated that the correlation functions ${\Cfull}_{\tilde u,\tilde g}(t,t',\bm k)$ can maintain stationarity on two two-dimensional slices in Fourier space, i.e.,
\beq
C_{\tilde u,\tilde g}^{\text{I}_x}(t,k_y,k_z)\equiv \int \frac{\text{d}k_x}{2\pi}\ {\Cfull}_{\tilde u,\tilde g}(t,t',\bm k),
\label{eqn:C2D}
\eeq
and
\beq
C_{\tilde u,\tilde g}^{xz}(t,k_x,k_z)\equiv {\Cfull}_{\tilde u,\tilde g}(t,t',k_x,k_y=0,k_z).
\label{eqn:C2D2}
\eeq
Since inverting $x\to -x$ or $y\to -y$ is equivalent to $S\to -S$ for the correlation functions, $C_{\tilde u,\tilde g}^{\text{I}_x,xz}$ will be invariant under $(k_x,k_y)\to(-k_x,-k_y)$, and so are the corresponding correlation times.
We can thus focus ourselves to half of the $(k_x,k_y)$ plane.
Furthermore, the quadratic functions $C_{\tilde u,\tilde g}^{\text{I}_x,xz}$ are also invariant under $k_z\to-k_z$.
We therefore only show the results in the $k_{x,y,z}\geq 0$ planes.

In the upcoming subsection, we present $\tau_{\tilde u,\tilde g}^{\text{I}_x}$ and their corresponding ratios on the $k_y$-$k_z$ plane. The second quantity (\ref{eqn:C2D2}) exclusively involves axisymmetric modes ($k_y=0$) and, as per the anti-dynamo theorem \citep{Cowling1933,Zel56}, cannot solely induce a dynamo.
For a comprehensive view, the $k_y=0$ slices [Equation~(\ref{eqn:C2D2})] and the energy densities for both velocity and helicity [corresponding to both quantities Equations~(\ref{eqn:C2D}) and (\ref{eqn:C2D2}) when $t=0$], are presented in the supplementary material.   

\subsection{Helicity probability distributions}
\begin{figure}
\centering
\includegraphics[width=\columnwidth]{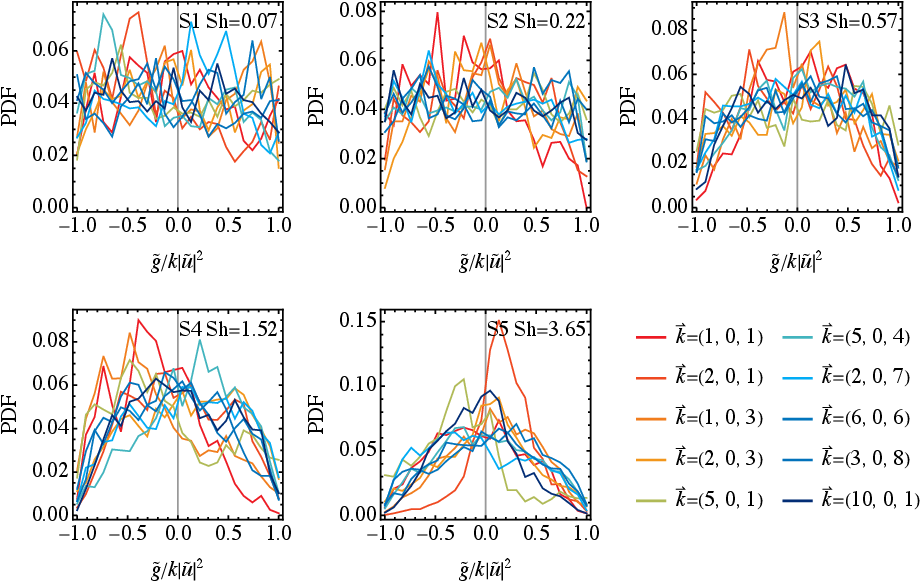}\\
\caption{PDFs of the normalized helicity $\tilde g/k|\tilde u|^2$ for a few modes for the shearing turbulence runs.}
\label{fig:S_pdf}
\end{figure}

In Figure~\ref{fig:S_pdf} we show $10$ typical PDF curves of the normalized helicity modes $\tilde g(\bm k)/k|\tilde u|^2$ for each of the runs.
At large enough shear rate (runs~S4 and S5) and low wave numbers ($k<\kf$), the PDFs are typically asymmetric with respect to zero,
possibly caused by the shear flow.
However, the asymmetry is different to those appear in the low-wave number modes in purely rotating cases (c.f. Figure~\ref{fig:R_pdf}).
For the rotating cases, helicity fluctuations appear to be small [i.e., the $\tilde g(\bm k)/k|\tilde u|^2$ values are close to zero], with a slight shift of the mean deviating from zero.
For the shearing cases, helicity fluctuations with one particular sign is more favored.
In this sense, low-wave number modes ($k<\kf$) in shearing turbulence would have stronger helicity fluctuations than those in the rotating turbulence.
At large wave numbers ($k>\kf$), the PDFs becomes symmetric about zero, and runs with lower shear rates (S1 and S2) exhibit flat PDFs, whereas at higher shear rates (S3 to S5), helicity fluctuations demonstrate less extreme values at $\tilde g/k|\tilde u|^2=\pm1$, suggesting a decrease in the occurrence of maximally helical fluctuations.
This is in contrast to the rotating cases (Figure~\ref{fig:R_pdf}) where high wave-number modes still exhibit flat PDFs at large rotation rates.

\subsection{Results of correlation times}
\begin{figure}
\centering
\includegraphics[width=\columnwidth]{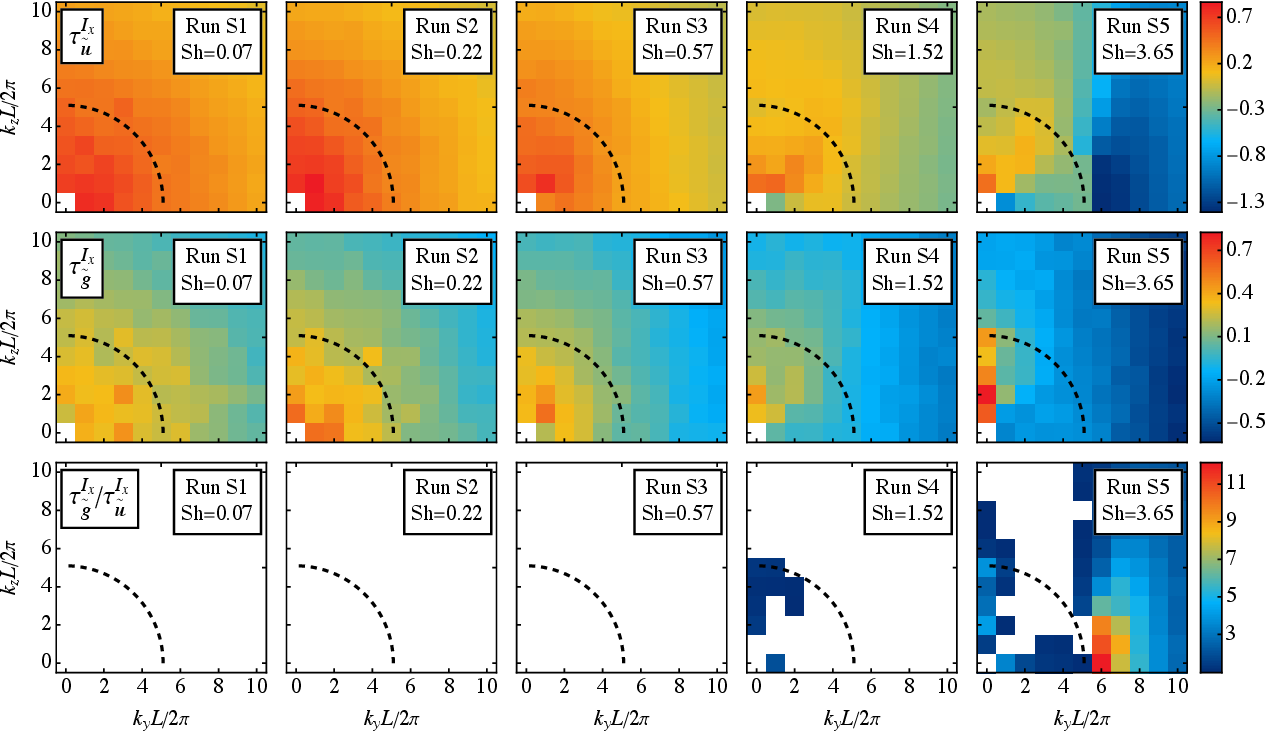}\\
\caption{For shearing turbulence, 
correlation times measured from $k_x$-integrated velocity and helicity correlations in the $(k_y,k_z)$ plane normalized by the eddy turnover time (upper and middle rows, $\log_{10}$ scale), and their ratios (bottom row, linear scale).}
\label{fig:S_Kyz}
\end{figure}

\begin{figure}
\centering
\includegraphics[width=0.5\columnwidth]{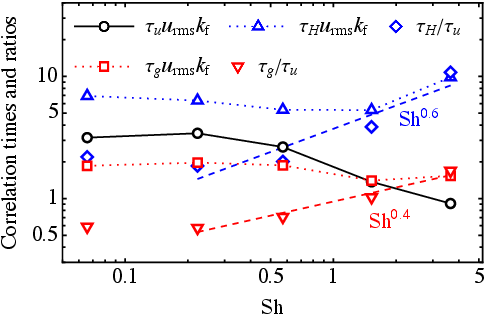}\\
\caption{For shearing turbulence, volume-averaged velocity and helicity correlation times and their ratios.
The dashed blue and dashed red lines are the fitted power-law relations for $\tau_H/\tau_u$ and $\tau_g/\tau_u$, respectively.
}
\label{fig:S0D}
\end{figure}

The correlation times for the $k_x$-integrated correlation functions [Equations~(\ref{eqn:C2D})] are presented in Figure~\ref{fig:S_Kyz}. Unlike cases of rotating turbulence, the instances where $\tau_{\tilde g}/\tau_{\tilde u}\geq 1$ are quite limited, with the exception of the most strongly sheared runs, S4 and S5.
This trend aligns with the red dashed line in Figure~\ref{fig:S0D}, illustrating the ratio between the volume-integrated correlation times $\tau_u$ and $\taug$. This ratio follows a $\Sh^{0.4}$ scaling and surpasses unity only around $\Sh\simeq 2$. Simultaneously, the ratio between the correlation time of the total helicity, $\tau_H$, and $\tau_u$ exhibits a similar $\Sh$-dependent scaling but with an overall amplitude exceeding unity. The absence of significant scale separation between velocity and helicity fluctuations at moderate shear can be ascribed to the elongation of vorticities along the shear direction in shearing flows. Additionally, the lack of distinctive vortex formation, which is observed in rotational flows, contributes to this phenomenon. 

The dynamo numbers $D_{H,g}$ [see \Eq{eqn:xi}] computed are less than $1.7\times 10^{-3}$, indicating that there is no negative diffussion in the shearing flows due to the helicity fluctuations. Considering the strength of helicity fluctuations and the lack of scale separation at moderate shear, we can conclude that, in the kinematic regime, the helicity fluctuations (incoherent $\alpha$ mechanism) may not be the possible driver of the shear dynamo simulations. 

\subsection{Shearing burgulence}
Burgers' equation is often used as a prototype equation to study turbulence, as it simplifies the Navier-Stokes equations by excluding the pressure gradient term. The form of Burgers' equation that we used to simulate the Burgulence counterpart of the shearing turbulence is given as:
\beq
\text{D}_t\bmu= Su_x{\bm e}_y-2\bm\Omega\times\bmu
+\frac{1}{\rho}\del\cdot\left(2\rho\nu\bm S\right)+\bm f,
\eeq
where $\bmu$ represents the velocity field, $S$ is the shear rate and $\nu$ is the kinematic viscosity. This equation captures essential features of turbulence, such as shock formation and energy dissipation, making it a useful tool for theoretical and numerical studies of turbulent flows.

Theories employing either the second-order-correlation approximation or the $\tau$ approximation closures in the shear-current effect argue for the essential role of the pressure gradient term in the Navier-Stokes equation \citep{SquireBhattacharjee2015pre,ZhouBlackman2021}. Without this term, the required turbulent transport coefficient becomes identically zero. Recent simulations demonstrating shear dynamos in burgulence \citep{Kapyla2021} seemingly challenge the shear-current effect as the primary driver in this context. However, this does not automatically confirm the incoherent $\alpha$ or helicity fluctuations as the correct explanation. The theoretical and numerical understanding of the role played by the pressure gradient term in the incoherent $\alpha$ effect is limited, and whether it persists in burgulence remains an open question. In this subsection, we present evidence that the helicity correlation weakens in burgulence, posing a challenge to the efficiency of the incoherent $\alpha$ effect in this scenario.

To avoid simulation crashes in burgulence runs, it was necessary to increase viscosity (refer to Table~\ref{tab:runs}).
This adjustment led to slightly lower Reynolds numbers compared to their counterparts in the Navier-Stokes simulations S1 to S5, but still comparable to those in \cite{Kapyla2021}.

\begin{figure*}
\centering
\includegraphics[width=\columnwidth]{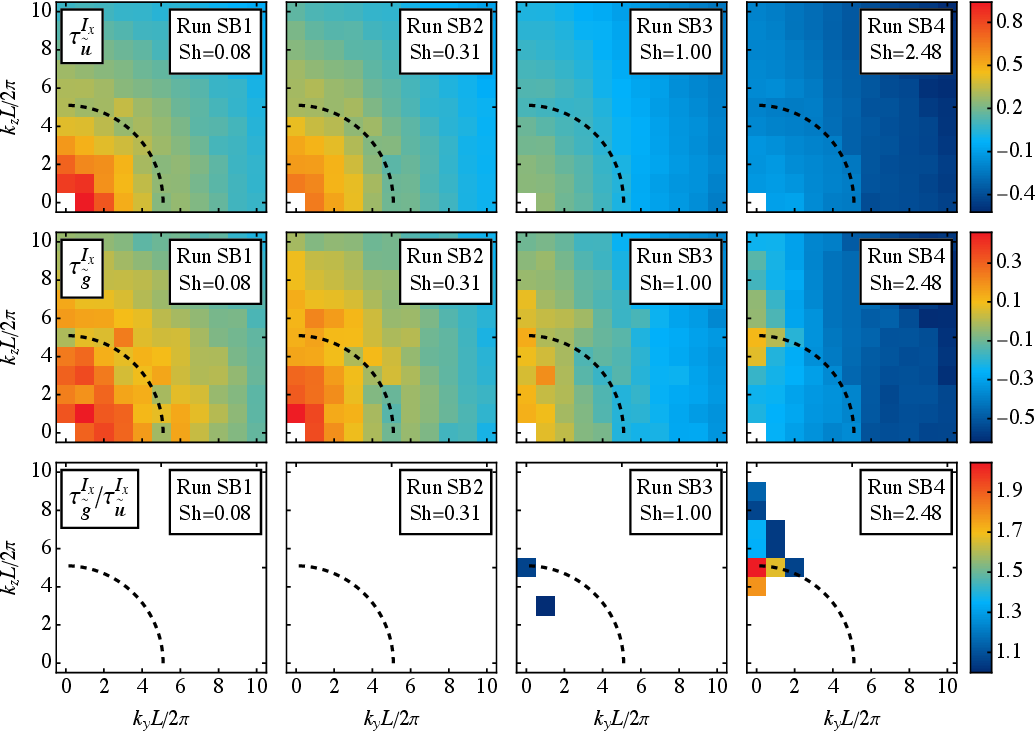}\\
\caption{For shearing burgulence,
correlation times measured from $k_x$-integrated velocity and helicity correlations in the $(k_y,k_z)$ plane and normalized by the eddy turnover time (upper and middle rows, $\log_{10}$ scale), and their ratios (bottom row, linear scale).}
\label{fig:SB_Kyz}
\end{figure*}

\begin{figure*}
\centering
\includegraphics[width=0.5\columnwidth]{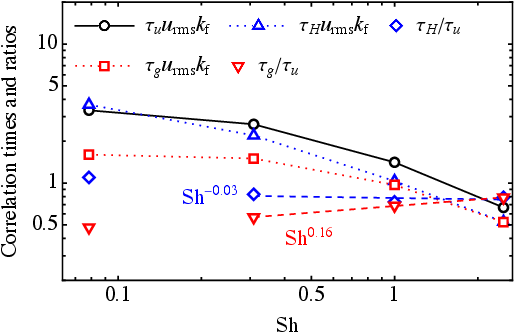}\\
\caption{Volume-averaged velocity and helicity correlation times and their ratios for the shearing burgulence cases.
The dashed blue and dashed red lines are the fitted power-law relations for $\tau_H/\tau_u$ and $\tau_g/\tau_u$, respectively.
}
\label{fig:ShearBurg_par}
\end{figure*}

The correlation times of velocity and helicity in the $k_y$-$k_z$ plane for shear-burgulence simulations are depicted in Figure~\ref{fig:SB_Kyz}.
Comparing the first two rows to their hydrodynamics counterparts in Figure~\ref{fig:S_Kyz},  we observe that both correlation times decrease more strongly with increasing $\Sh$ in the case of burgulence, especially in the low wave number ($k<\kf$) region.
The resulting ratio of the time scales (the third row) shows no substantial scale separation, similar to the shearing hydrodynamic turbulence.
The correlation times of volume-averaged correlation functions are presented in Figure~\ref{fig:ShearBurg_par}.
With increasing $\Sh$, $\tau_u$ decreases at a rate similar to the hydrodynamic case, see Figure~\ref{fig:S0D}. However, both helicity correlation times $\tau_{H,g}$ are suppressed more strongly, resulting in lower magnitudes. This leads to much smaller time scale ratios $\tau_{H,g}/\tau_u$ at $\Sh \gtrsim 1$.

\subsection{Keplerian runs}
For Keplerian runs, we choose the value $q=S/\Omega=3/2$ resulting in Coriolis number $\Co=4\Sh/3$.
This case is more representative of astrophysical scenarios, such as in galaxies and accretion disks, where $q$ can vary from $1$ to $2$.
The methodology applied for these runs remains consistent with the shear runs described in the previous subsection.

\begin{figure*}
\centering
\includegraphics[width=\columnwidth]{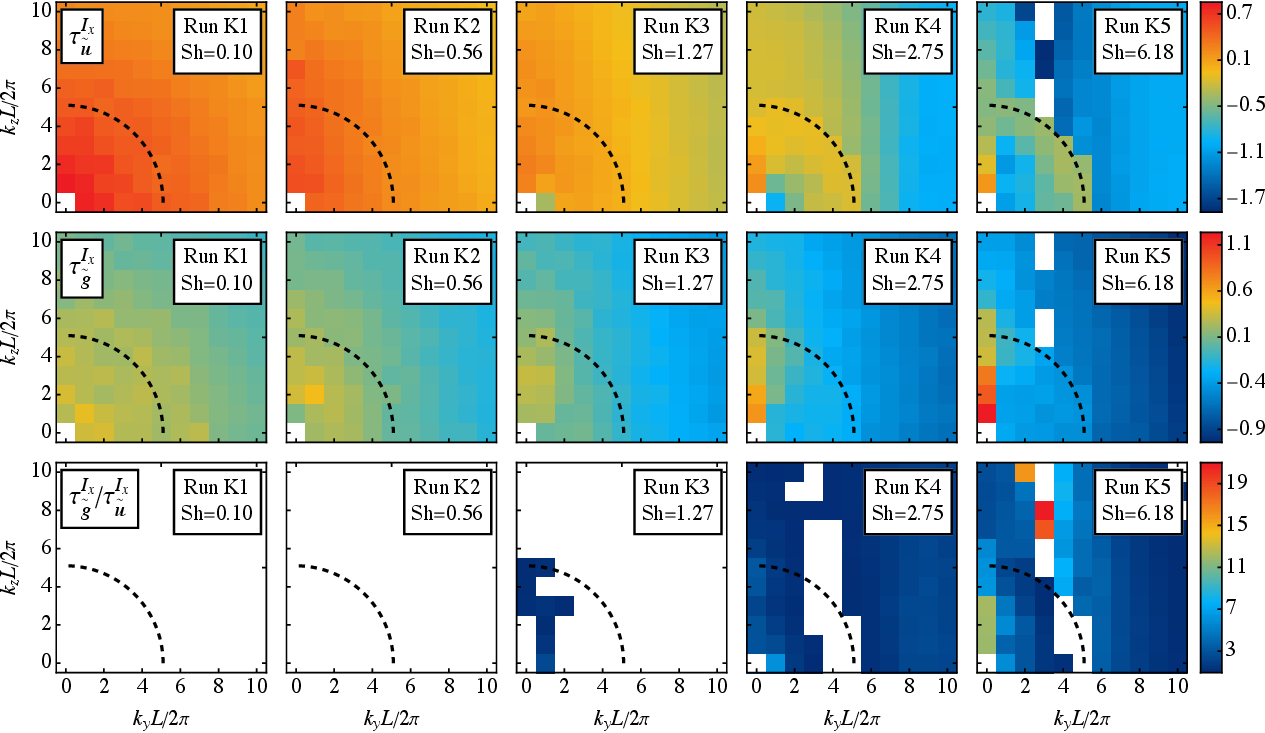}\\
\caption{For Keplerian turbulence, 
correlation times measured from $k_x$-integrated velocity and helicity correlations in the $(k_y,k_z)$ plane and normalized by the eddy turnover time (upper and middle rows, $\log_{10}$ scale), and their ratios (bottom row, linear scale).}
\label{fig:K_Kyz}
\end{figure*}

\begin{figure*}
\centering
\includegraphics[width=0.5\columnwidth]{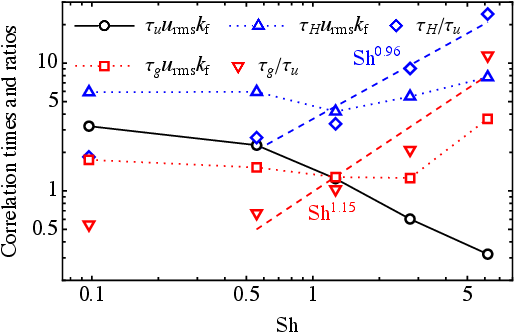}\\
\caption{Volume-averaged velocity and helicity correlation times and their ratios for the Keplerian turbulence cases.
The dashed blue and dashed red lines are the fitted power-law relations for $\tau_H/\tau_u$ and $\tau_g/\tau_u$, respectively.}
\label{fig:Kep_par}
\end{figure*}

In \Fig{fig:K_Kyz}, Fourier space results reveal a more distinct scale separation compared to both the shearing turbulence (\Fig{fig:S_Kyz}) and shearing burgulence (Figure~\ref{fig:SB_Kyz}) cases.
This enhanced separation is attributed to the additional rotation, known to form condensate through inverse cascade producing large-scale vortices.
This time-scale separation persists despite the elongation effect of the vortices in the direction of the shear. 
The volume-integrated time scales shown in \Fig{fig:Kep_par} also display a larger scale separation:
The magnitudes of both ratios, $\tau_{H,g}/\tau_u$ are similar to the shearing cases [c.f. Figure~\ref{fig:S0D}] at $\Sh=0.5$. However, their scalings with $\Sh$ become much stronger, with $\tau_H/\tau_u\propto \Sh^{0.96}$ and $\tau_g/\tau_u\propto \Sh^{1.15}$.
Interestingly, these power-law indices are larger than either purely rotating ($\propto\Cokf^{0.85}$ and $\propto\Cokf^{0.09}$, respectively) or purely shearing (with weak counter-rotation, $\propto\Sh^{0.63}$ and  $\propto\Sh^{0.38}$, respectively) cases, revealing a constructive collaboration of the rotation and the shear flow. 

In these Keplerian runs, we also find that the dynamo number $D_{H,g}$ [see \Eq{eqn:xi}] is less than $1.5\times 10^{-3}$, suggesting that the strength of helicity fluctuations is relatively weak. It is important to note that this strength depends on the Mach number of the flows, which in our simulations is around $0.1$. The strength of these helicity fluctuations could increase with higher Mach numbers. Therefore, further exploration of this effectiveness on the dynamo mechanism in different flow regimes, such as transonic and supersonic flows more relevant to astrophysical scenarios, would be of significant interest. However, such an investigation is beyond the scope of the current work.

\section{Comparison with previous shear dynamo simulations}
\label{sec:compare}

In the previous sections we have obtained the degree of time-scale separation between helicity and velocity fluctuations in rotating, shearing, and Keplerian hydrodynamic turbulence, as well as in shearing burgulence.
In this section we compare our parameter space with those in previous numerical simulations of large-scale dynamos in shear flows and seek for implications in the light of Figures~\ref{fig:R0D}, \ref{fig:S0D}, \ref{fig:ShearBurg_par}, and \ref{fig:Kep_par}.

\begin{figure}
\centering
\includegraphics[width=0.8\columnwidth]{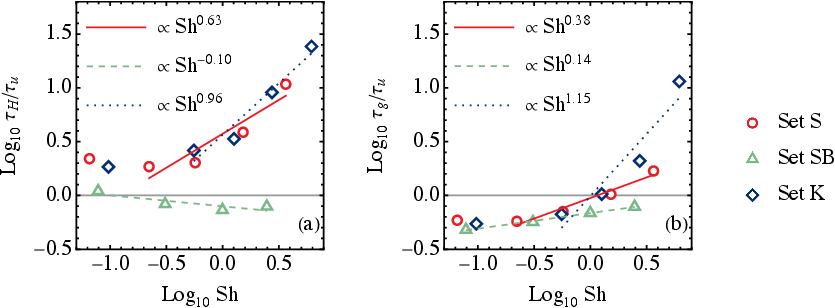}\\
\caption{Ratios of time scales $\tau_{H,g}/\tau_u$ for different sets of runs:
Shearing turbulence  (S), shearing burgulence (SB),
and Keplerian turbulence (K).
Datasets are the same as those in Figures~\ref{fig:S0D}, \ref{fig:ShearBurg_par}, and \ref{fig:Kep_par}.}
\label{fig:all_ratios}
\end{figure}

The general idea we gained from previous section is that the time scale separation between helicity and velocity fluctuations increases with increasing rotation or shear rate.
This increase is quantified using $\zeta=\tau_{H,g}/\tau_u$, and we present all the results of $\zeta$ for the runs with shear in Figure~\ref{fig:all_ratios} to facilitate comparison. 
The renovating-flow model in \cite{JingadeSingh2021} suggests that there exists a critical value $\zeta_\text{crit}$ ($=3$ in their work) above which the incoherent $\alpha$ effect is capable of driving a shear dynamo.
The exact value of $\zeta_\text{crit}$ in hydrodynamical or magnetohydrodynamical turbulence might be different from $3$, given the simplifications made in the model,
but it is reasonable to expect the existence of some $\zeta_\text{crit}>1$, since the coherent effect has to be strong enough to overcome turbulent dissipation.
Combining that (i) $\zeta$ increases with increasing $\Sh$ and (ii) $\zeta_\text{crit}>1$,
it follows that the incoherent $\alpha$ effect requires a critical value of $\Sh$.

A critical shear rate is also predicted in the alternative explanation of shear dynamos, the shear-current effect, but there is an important distinction regarding its scale-dependence.
The shear-current effect model predicts the dynamo growth rate for the $xy$-averaged field to be \citep[see, e.g.,][]{ZhouBlackman2021}
\beq
\gamma=\left[\frac{(\beta_{xx}-\beta_{yy})^2}{4}k_1^4
+\beta_{xy}\beta_{yx}k_1^4+S\beta_{yx}k_1^2\right]^{1/2}
-\frac{\beta_{xx}+\beta_{yy}}{2}k_1^2,
\eeq
where $\beta_{ij}$ is the magnetic turbulent diffusivity tensor (which depends on $\Sh$),
and $k_1$ is the wave number of the large-scale magnetic field of interest.
Considering the simplified case where $\beta_{xx}=\beta_{yy}=\beta_0$ and $\beta_{xy}=0$,
we have
\beq
\gamma=\sqrt{S\beta_{yx}k_1^2}-\beta_0k_1^2
=-\beta_0\left(k_1-\frac{\sqrt{S\beta_{yx}}}{2\beta_0}\right)^2
+\frac{S\beta_{yx}}{4\beta_0}.
\label{eqn:gamma_sce}
\eeq
The first equality implies that a minimal shear rate is required for the growth rate to be positive.
However, the second equality shows that the dynamo growth rate remains finite for any small $\Sh$, as long as the mode with the fastest growth rate fits into the simulation box. This contrasts with the incoherent $\alpha$ effect, which requires a critical shear rate $\Sh_{\text{crit}}$ for maximum growth, even when $k_1$ corresponds to the fastest growing mode. This distinction can help differentiate between these two dynamo mechanisms.

\cite{Yousef2008} investigated the shear dynamo problem in the kinematic phase using tall simulation boxes, low shear rates, and without rotation. In their study, the vertical scales of the simulation domain are always larger than the typical scales of the large-scale magnetic fields, allowing the fastest growing mode to be captured. They used a dimensionless shear parameter $\Sh \lesssim 1/(3\pi) \simeq 0.1$ based on our definition.
All simulations in their study exhibited shear dynamos, even at very low values of $\Sh$ \citep[notably, for $\Sh \gtrsim 0.05$, the large-scale fields are disrupted in the nonlinear stage; see][]{TeedProctor2016}. Therefore, combining our results on the threshold shear needed to bring in the time-scale separation with the findings from \cite{Yousef2008}, it suggests that the observed shear dynamos are unlikely to be driven by the incoherent $\alpha$ effect.

It is also worth noting that both \cite{Yousef2008} and \cite{TeedProctor2016} used $\ReN=\ReM\simeq5$, which is sub-critical for a small-scale dynamo. Hence, their simulations fall outside the realm of the magnetic shear-current effect, which requires strong magnetic fluctuations \citep{SquireBhattacharjee2015pre}. Furthermore, the kinetic shear-current effect \citep{RogachevskiiKleeorin2003} is also likely ruled out because the steep kinetic energy spectrum at such low $\ReN$ is unfavorable for the shear dynamo \citep{ZhouBlackman2021}. Given that the current study also disfavors the incoherent $\alpha$ driver, it becomes puzzling what drives large-scale dynamos in non-helical shearing turbulence when both Reynolds and magnetic Reynolds numbers are not very large. This suggests a need for further investigation into the properties of shear dynamos, potentially leading to a new classification based on these properties.

As for large-scale dynamos in shearing burgulence, \cite{Kapyla2020} explored cases with $\Sh=1.6$ and $(\ReN,\ReM)\simeq(0.5,2)$ and $(\ReN,\ReM)\simeq(0.7,12)$. By using the nonlinear test-field method, they ruled out the shear-current effect as the driver of the shear dynamo. In the current study, we find that the time scale separation between the velocity and helicity fluctuations is strongly suppressed in shearing burgulence compared to its hydrodynamic counterpart (i.e., the S and SB runs in Figure~\ref{fig:all_ratios}). This result seemingly suggests the incapability of the incoherent $\alpha$ effect with Burgers' equation. However, we cannot rule out the possibility that the critical ratio between time scales $\tau_{H,g}/\tau_u$ needed for a shear dynamo in shearing burgulence is lower than that in the hydrodynamic case.

The consideration of shearing burgulence also raises the important question of the Mach number dependence of $\zeta$ in rotating or shearing flows. The current study only explored weakly compressible hydrodynamical turbulence at $\Ma\simeq 0.1$. For turbulence in accretion disks, the Mach number depends on the Shakura-Sunyaev viscosity parameter $\alpha_\text{SS}$, with $\Ma\simeq\alpha_\text{SS}^{1/2}$ \citep{Blackman1998}. In moderately compressible flows, helicity fluctuations might become stronger, potentially allowing the dynamo number to cross the threshold of the incoherent $\alpha$ effect. Investigating how time-scale separation and helicity fluctuation amplitudes vary with $\Ma$ could provide valuable insights into the dynamo problem in accretion disks.

\section{Conclusions}
\label{sec:conclusion}

In this investigation, we scrutinized the correlation times of velocity and helicity fluctuations in diverse fluid flow scenarios, including rotating flows, shearing flows, Keplerian flows, and shearing burgulence.
Our primary objective is to gauge the correlation time between velocity and helicity fluctuations, crucial for elucidating \revise{}{the kinematic} dynamo mechanisms in nonhelical turbulent flows. While both the shear-current effect and helicity fluctuations are leading candidates for explaining shear dynamos, we focused specifically on the latter in this study. 

We have investigated the correlation times for individual Fourier modes as well as for volume-averaged values, each used in different theories.
In the context of rotating flows, a notable temporal separation between velocity and helicity fluctuations is revealed, particularly prominent in large-scale modes influenced by the Coriolis force arising from rotation.
In Fourier space (Figure~\ref{fig:R2D}), an increase in the rotation rate results in an overall elevation of the magnitudes of time scale ratios and a greater number of modes where this ratio surpasses unity.
These modes are preferentially situated along the rotation axis, particularly noticeable in smaller scales with wave numbers close to, but surpassing, the forcing wave number.
On the other hand, the magnitudes of both velocity and helicity correlation time scales decreases when the rotation rate increases.
This prolonged persistence of helicity fluctuations relative to velocity fluctuations is ascribed to the formation of cyclonic and anti-cyclonic vortices within the rotating turbulence.

In shearing cases (with weak negative rotation to suppress the vorticity dynamo), non-axisymmetric modes conducive to dynamos exhibited more coherent helicity fluctuations only in strongly sheared flows ($\Sh\simeq 2$, Figure~\ref{fig:S_Kyz}).
The lack of required scale separation for helicity and velocity fluctuations in non-axisymmetric modes that are crucial for dynamo initiation was linked to vortex elongation along the shear direction.
Therefore, we infer that helicity fluctuations may not be the driving force for shearing dynamos in non-rotating shearing flows, necessitating a re-evaluation or further categorization of shear dynamos based on their properties.
For Keplerian flows, time scale separation in non-axisymmetric modes is observed at moderate shear rates ($\Sh\gtrsim1$, Figure~\ref{fig:K_Kyz}), which suggests that helicity fluctuations may play a more significant role since the shear-current effect is suppressed \citep{ZhouBlackman2021}.

Furthermore, the shear-current effect is estimated to vanish in shearing burgulence \citep{SquireBhattacharjee2016,ZhouBlackman2021}. Our results imply that removing the pressure gradient in the Navier-Stokes equation introduces a decoherence effect to helicity, challenging the incoherent $\alpha$ effect (Figure~\ref{fig:SB_Kyz}). Consequently, the driver for large-scale dynamos in shearing burgulence remains elusive.

An essential consideration for dynamos is the strength of helicity fluctuations. In our study, conducted for subsonic flows with Mach numbers $\simeq0.1$, the normalized ratio of helicity fluctuations to velocity fluctuations, as expressed in Equation (\ref{eqn:xi}), remained below $2\times 10^{-3}$. Such weak fluctuations raise doubts about triggering dynamos, even in Keplerian runs. Considering the dependence on the fourth power of velocity in the numerator and quadratic power in the denominator, an extension of this study to transonic or supersonic cases with higher Mach numbers could significantly increase fluctuation strength, potentially leading to large-scale dynamos. This extension may hold implications for accretion disk dynamos.

\backsection[Acknowledgements]{
We thank the anonymous referees for their constructive suggestions which help improve the paper.
We thank Axel Brandenburg, Jennifer Schober, and Alberto Roper Pol for their help in developing the numerical methods.
We also thank Axel Brandenburg, Eric G. Blackman, and Matthias Rheinhardt for insightful discussions.
Nordita is partially supported by Nordforsk.
We acknowledge the allocation of computing resources provided by the Swedish National Allocations Committee at the Center for Parallel Computers at the Royal Institute of Technology in Stockholm and Link\"oping.
Part of the numerical simulations were carried out on the Siyuan Mark-I and the ARM platform clusters supported by the Center for High Performance Computing at Shanghai Jiao Tong University,
and the Astro cluster supported by Tsung-Dao Lee Institute.
NJ acknowledges the use of the High
Performance Computing (HPC) resources made available by
the Computer Centre of the Indian Institute of Astrophysics
(IIA).} 

\backsection[Funding]{HZ acknowledges support from grant number 2023M732251 from the China Postdoctoral Science Foundation.}

\backsection[Data availability statement]{The data that support the findings of this study will be shared on reasonable request to the authors.}

\backsection[Author ORCIDs]{H. Zhou, https://orcid.org/0000-0002-2991-5306;
N. Jingade, https://orcid.org/0000-0002-8079-9566}

\backsection[Declaration of Interests]{The authors report no conflict of interest.}

\appendix

\section{Time-stationarity in Fourier space}
\label{sec:appx1}
Consider a Cartesian frame with coordinates $(X,Y,Z)$, where the shear flow is denoted by $\bm U^\text{shear}=SX{\bm e_Y}$. In the lab frame, the most general form of a two-point unequal-time auto-correlation function can be derived using Galilean invariance \citep{SridharSingh2014}. This invariance is associated with measurements made by observers whose velocity, relative to the lab frame $(t,\bfX)$ matches that of the background shear flow. These comoving observers can be characterized by $\bfxi=(\xi_1,\xi_2,\xi_3)$, representing the position of the origin of the comoving observer at the initial time zero. The position of the origin of the comoving observer at time $t$ is given by
\beq
\bfX_c(t,\bfxi)  = (\xi_1, \xi_2+St\xi_1, \xi_3)
\eeq
Since the statistics of the velocity field remains the same, the velocity correlator measured in the lab frame or the corresponding comoving frame must be the same,
and can be expressed as  
\beq
\bar{\bmu^\text{lab}(t,\bfX)\cdot \bmu^\text{lab}(t',\bfX')}
= \bar{\bmu^\text{lab}(t,\bfX + \bfX_c(\bfxi,t))\cdot \bmu^\text{lab}(t',\bfX'+ \bfX_c(\bfxi,t'))}
\eeq
The Navier-Stokes equation does not explicitly depend on time, indicating that the velocity statistics must be time-stationary. Consequently, a Galilean-invariant velocity correlator with an arbitrary time shift $t_0$ can be expressed as
\begin{align}
&\bar{\bmu^\text{lab}(t,\bfX)\cdot \bmu^\text{lab}(t',\bfX')}\notag\\
=&\bar{\bmu^\text{lab}(t+t_0,\bfX + \bfX_c(\bfxi,t+t_0))\cdot \bmu^\text{lab}(t'+t_0,\bfX'+ \bfX_c(\bfxi,t'+t_0))}.
\end{align}
This equation holds for any arbitrary $\bfxi$ and time $t_0$. By choosing $\bfxi = -\bfX'$ and $t_0$ as $-t'$, we obtain the most general form of the correlator as follows
\beq
\bar{\bmu^\text{lab}(t,\bm X)\cdot \bmu^\text{lab}(t',\bm X')}
={\Cfull}_{u,\text{lab}}(\bm X-\bm X'-S(t-t')X{\bm e_Y},t-t').
\label{eqn:Culab}
\eeq
While ${\Cfull}_{u,\text{lab}}$ is time-stationary, it explicitly depends on $X$, therefore inhomogeneous in space. This characteristic is akin to the velocity correlator that would result from applying the Taylor hypothesis to frozen turbulence.

In a shearing-frame transformation, a set of Cartesian coordinates $(x,y,z)$ whose directions are parallel to those of the lab frame is linked to the lab-frame coordinates by
\beq
\bm x=\bm X-StX{\bm e_Y},
\label{eqn:Sheartrans}
\eeq
and the velocity field in the shearing frame is defined by
\beq
\bmu(t,\bm x)=\bmu(t,\bfX-St X{\bm e_Y})=\bmu^\text{lab}(t,\bm X).
\eeq
Applying the transformation to Equation~(\ref{eqn:Culab}) yields
\beq
\bar{\bmu(t,\bmx)\cdot \bmu(t',\bmx')}={\Cfull}_{u}(\bmx-\bmx'+St'(x-x'){\bm e_Y},t-t'),
\label{eqn:Cu_shear_frame}
\eeq
which restores spatial homogeneity at the cost of losing time stationarity property when $x\neq x'$. 

To measure the correlation time in Fourier space, we require the velocity correlator to be both spatially homogeneous and time-stationary. Thus, we aim to construct a correlator that preserves both of these properties. With the shearing transformation for the spatial coordinate, we can derive the corresponding transformation for its Fourier counterpart. This transformation is obtained from the conservation of the phase of the wave, $\bfK\cendot\bfX = \bfk\cendot\bfx$ \citep[see Appendix in][for details]{SridharSingh2010}, where $\bfK$ is the lab frame wavevector and $\bfk$ is the shearing frame wavevector. Using Equation (\ref{eqn:Sheartrans}), we find $\bfK= \bfk - S t k_y \hat{\bfX}$.
 
 The spatial homogeneity of Equation~(\ref{eqn:Cu_shear_frame}) ensures $\bar{\tilde \bmu^*(t,\bm k)\cdot\tilde \bmu(t',\bm k')}=0$ unless $\bm k =\bm k'$. Consequently, we can define
\beq
{\Cfull}_{\tilde u}(t,t',\bm k)=\bar{\tilde \bmu^*(t,\bm k)\cdot\tilde \bmu(t',\bm k)},
\eeq
whose general form is given by Fourier transforming Equation~(\ref{eqn:Cu_shear_frame}),
\begin{align}
{\Cfull}_{\tilde u}(t,t',\bm k)=&\int\text{d}^3r\ \ e^{-i\bm k\cdot\bm r}{\Cfull}_u({\bm r} +St' r_x{\bm e_Y},t-t');\quad \mbox{where}\quad {\bm r} = \bfx-\bfx'\notag\\
=&\widetilde{{\Cfull}_u}(\bfk-St'k_y \hat{\bfX},t-t').
\label{eqn:fourierC}
\end{align}
Note that the Fourier space correlator $C_{\tilde u}(t,t',\bm k)$ is not time-stationary, given its explicit dependence on the time $t'$. The time-stationary property is crucial for building statistics over time, facilitating the averaging required to obtain the correlator. Consequently, we opt for specific correlators constructed from \Eq{eqn:fourierC}, designed to preserve the time-stationary property.

Firstly if we restrict ourselves to the $k_y=0$ plane, we obtain a set of correlator that are time-stationary, denoted as
\beq
C_{\tilde u}^{xz}(t,t',k_x,k_z)={\Cfull}_{\tilde u}(t,t',k_x,k_y=0,k_z)
=\widetilde{{\Cfull}_u}(k_x,0,k_z,t-t').
\label{eqn:Cu2D2}
\eeq
The other set is obtained by integrating the correlator \Eq{eqn:fourierC}, over $k_x$. By applying the transformation property of Fourier wave vectors, the integrals of the correlators can be expressed as
\beq
\int\text{d}k_x \text{d}k_y\ \widetilde{{\Cfull}_u}(k_x-St'k_y,k_y,k_z,t-t')
= \int\text{d}K_x \text{d}K_y\ \widetilde{{\Cfull}_u}(K_x,K_y,k_z,t-t') 
\eeq
where $K_x = k_x-St'k_y$ and $K_y = k_y$ and the Jacobian of the transformation between lab frame and shearing frame wavevectors is unity. Defining a correlator by integrating over the $K_x$ wavevector, 
\beq
C_{\tilde u}^{\text{I}_x}(t,t',k_y,k_z)
=\int\frac{\text{d}K_x}{2\pi}\ \widetilde{{\Cfull}_u}(K_x,k_y,k_z,t-t')
\label{eqn:Ctu2D2}
\eeq
we obtain the time-stationary correlation function, as they depend only on the difference in time $(t-t')$. 

We can express the integrated correlator \Eq{eqn:Ctu2D2} in terms of the velocity as
\begin{align}
C_{\tilde u}^{\text{I}_x}(t,t',k_y,k_z) = &\int\, \frac{\ud k_x}{2\pi}\, \bar{\widetilde{\bfu^*}(k_x,k_y,k_z,t)\cendot\widetilde{\bfu}(k_x,k_y,k_z,t')}\notag\\
= &\int \ud x\, \bar{\bfu^*(x,k_y,k_z,t)\cendot\bfu(x,k_y,k_z,t')}
\end{align}
where the last equality is due to the application of Parseval's theorem in one dimension.
In our numerical experiments, we approximate the last expression by integrating $x$ in the range $-5\pi/32\leq x\leq5\pi/32$, which captures the small-scale modes responsible for the dynamo transport coefficients.
The compromise of integrating over a limited range of $x$ is made because the two-time correlators have to be calculated from the Fourier modes during the post-processing, and the output modes are limited due to the high cadence needed to capture short correlation times. 

It's crucial to highlight that \Eq{eqn:Cu2D2} provides the axisymmetric modes, while \Eq{eqn:Ctu2D2} yields the non-axisymmetric modes in an integrated form along one dimension. The significance lies in the fact that only the non-axisymmetric modes contribute to dynamos, making them particularly relevant for dynamo studies.


\bibliographystyle{jfm}
\bibliography{refs}

\end{document}